\documentclass[useAMS,usenatbib,usegraphicx]{mn2e}
\usepackage{multirow}\topmargin=-1.7cm
\title[Orochi]{Detection of an ultra-bright submillimeter galaxy in the Subaru/XMM-Newton Deep Field using AzTEC/ASTE}
\author[S. Ikarashi et al.]{S. Ikarashi$^{1}$\thanks{E-mail: ikarashi@ioa.s.u-tokyo.ac.jp}, K. Kohno$^{1, 2}$, J. E. Aguirre$^{3}$, I. Aretxaga$^{4}$,  V. Arumugam$^{5}$,\newauthor
J. E. Austermann$^{6}$,  J. J. Bock$^{7, 8}$, C. M. Bradford$^{7, 8}$, M. Cirasuolo$^{5, 9}$ L. Earle$^{10}$,\newauthor
 H. Ezawa$^{11}$, H. Furusawa$^{13}$, J. Furusawa$^{13}$, J. Glenn$^{10}$, B. Hatsukade$^{12}$, D. H. Hughes$^{4}$, \newauthor
 D. Iono$^{12}$, R. J. Ivison$^{5, 9}$ S. Johnson$^{14}$, J. Kamenetzky$^{10}$, R. Kawabe$^{12}$, R. Lupu$^{3}$,\newauthor
   P. Maloney$^{10}$, H. Matsuhara$^{15}$, P. D. Mauskopf$^{16}$, K. Motohara$^{1}$,  E. J. Murphy$^{8}$, \newauthor
    K. Nakajima$^{17}$, K. Nakanishi$^{11}$, B. J. Naylor$^{7}$, H. T. Nguyen$^{7}$,  T. A. Perera$^{18}$, \newauthor
     K. S.  Scott$^{3}$, K. Shimasaku$^{2, 17}$, T. Takagi$^{15}$, T. Takata$^{13}$, Y. Tamura$^{12}$, K. Tanaka$^{19}$,\newauthor
     T. Tsukagoshi$^{1}$, D. J. Wilner$^{20}$, G. W. Wilson$^{14}$, M. S.  Yun$^{14}$, J. Zmuidzinas$^{7, 8}$
\\
$^{1}$Institute of Astronomy, University of Tokyo, 2-21-1 Osawa, Mitaka, Tokyo 181-0015, Japan\\
$^{2}$Research Center for the Early Universe, School of Science, University of Tokyo, 7-3-1 Hongo, Bunkyo, Tokyo 113-0033, Japan\\
$^{3}$Department of Physics and Astronomy, University of Pennsylvania, Philadelphia, PA 19104, USA \\
$^{4}$Instituto Nacional de Astrof\'{\i}sica, \'Optica y Electr\'onica (INAOE), Aptdo. Postal 51 y 216, 72000 Puebla, Pue., Mexico  \\
$^{5}$Institute for Astronomy, University of Edinburgh, Royal Observatory, Blackford Hill, Edinburgh EH9 3HJ, UK \\
$^{6}$Center for Astrophysics and Space Astronomy, University of Colorado, Boulder, CO 80309, USA\\
$^{7}$Jet Propulsion Laboratory, Pasadena, CA 91109, USA \\
$^{8}$California Institute of Technology, Pasadena, CA 91125, USA \\
$^{9}$UK Astronomy Technology Centre, Royal Observatory, Blackford Hill, Edinburgh EH9 3HJ, UK\\
$^{10}$Dept. of Astrophysical and Planetary Sciences, University of Colorado, 389-UCB, Boulder, CO 80309, USA\\
$^{11}$ALMA Project Office, National Astronomical Observatory, 2-21-1 Osawa, Mitaka, Tokyo 181-8588, Japan\\
$^{12}$Nobeyama Radio Observatory, Minamimaki, Minamisaku, Nagano 384-1305, Japan \\
$^{13}$Astronomy Data Center, National Astronomical Observatory, Mitaka, Tokyo 181-8588, Japan \\
$^{14}$Department of Astronomy, University of Massachusetts, Amherst, MA 01003, USA\\
$^{15}$Institute of Space and Astronautical Science, Sagamihara, Kanagawa 229-8510, Japan\\
$^{16}$School of Physics \& Astronomy, Cardiff University, Queens Buildings, The Parade, Cardiff CF24 3AA, UK \\
$^{17}$Department of Astronomy, University of Tokyo, Hongo 7-3-1, Bunkyo-ku, Tokyo 113-0033, Japan \\
$^{18}$Illinois Wesleyan University 1312 Park Street Bloomington, IL 61701, USA \\
$^{19}$Department of Physics, Faculty of Science and Technology, Keio University, 3-14-1 Hiyoshi, Kohoku-ku, Yokohama, Kanagawa 223-8522, Japan \\
$^{20}$Harvard-Smithsonian Center for Astrophysics, 60 Garden Street, Cambridge, MA 02138, USA \\
}
\begin{document}
\date{Accepted ... ; Received ... ; in original form 2010 September 07}
\pagerange{\pageref{firstpage}--\pageref{lastpage}} \pubyear{2010}
\maketitle

\label{firstpage}
\begin{abstract}
We report the detection of 
an extremely bright ($\sim$37 mJy at 1100 $\mu$m and $\sim$91 mJy at 880 $\mu$m) submillimeter galaxy (SMG),
 AzTEC-ASTE-SXDF1100.001 (hereafter referred to as SXDF1100.001 or Orochi), discovered in 1100 $\mu$m observations of the Subaru/XMM-Newton Deep Field using AzTEC on ASTE. 
Subsequent CARMA 1300 $\mu$m and SMA 880 $\mu$m observations successfully
pinpoint the location of Orochi and suggest that it has two components,  
one extended  (FWHM of $\sim$ 4$^{\prime\prime}$) and one compact (unresolved).
Z-Spec on CSO has also been used to obtain a wide band spectrum from 190 to 308 GHz, although no significant emission/absorption lines are found. The derived upper limit to the line-to-continuum flux ratio is 0.1--0.3 (2 $\sigma$) across the Z-Spec band.

Based on the analysis of the derived spectral energy distribution  from optical to radio wavelengths of possible counterparts near the SMA/CARMA peak position, we suggest that Orochi is a lensed, optically dark SMG lying at $z \sim 3.4$ behind a foreground, optically visible (but red) galaxy at $z \sim 1.4$.
The deduced apparent (i.e., no correction for magnification) infrared luminosity ($L_{\rm IR}$) and star formation rate (SFR) are $6 \times 10^{13}$ $L_{\odot}$ and 11000 $M_{\odot}$ yr$^{-1}$, respectively, assuming that the $L_{\rm IR}$ is dominated by star formation. These values suggest that Orochi will consume its gas reservoir within a short time scale ($3 \times 10^{7}$ yr), which is indeed comparable to those in extreme starbursts like the centres of local ULIRGs.

\end{abstract}


\section{Introduction}

Recent rapid advancements in wide and deep surveys at millimeter/submillimeter (mm/submm) wavelengths
have led to successive discoveries of numerous  mm/submm-bright galaxies 
(SMGs) in the early universe. 

After the pioneering works performed with the SCUBA \citep{b47,b46} on the JCMT 15 m telescope \citep[e.g.,][]{b48,b49,b105,b50},  
numerous mapping surveys at $\lambda=$ 850 $\mu$m have been conducted toward blank fields
\citep[e.g.,][]{b51,b52,b53} and over-dense regions \citep[e.g.,][]{b54,b106}.
MAMBO \citep{b55} on the IRAM 30 m telescope 
and BOLOCAM \citep{b56} on the CSO 10 m have also 
been used to obtain 1200 $\mu$m/1100 $\mu$m images 
of blank fields \citep[Greve et al. 2004, 2008;][]{b60,b56}
and over-dense regions \citep[e.g.,][]{b59}.

A new mm-wavelength bolometer camera, AzTEC for 1100 $\mu$m \citep{b4}, was mounted on JCMT and produced wide-area ($\sim$ a few 100-1000 arcmin$^2$ scale) images of well studied fields such as COSMOS \citep{b6}, GOODS-N \citep{b62}, Lockman hole, SXDF \citep{b63}, and over-dense regions \citep{b109}.

New submm wave telescopes in northern Chile, i.e., ASTE 10 m and APEX 12 m telescopes, are now also equipped with bolometer cameras, i.e., AzTEC for 1100 $\mu$m and LABOCA for 870 $\mu$m \citep{b61}. Owing to the very suitable atmospheric conditions of the site, these telescopes routinely obtain wide-area images of various fields such as
SSA22 \citep{b64}, 
ECDF-S/GOODS-S \citep{b65,b84}, ADF-S \citep{b107}, 
a proto-cluster \citep{b66} and a cluster \citep{b5}.
A wider area ($\sim$10 deg$^2$ scale) short submm survey has been conducted with  BLAST \citep{b85}, and $\sim$100 deg$^2$ scale mm/submm surveys are now coming using SPT \citep{b71} and Herschel \citep{b82,b86}.

One of the important findings from these recent mm/submm surveys is the detection of ultra-bright populations of SMGs
and it is predicted that this ultra-bright population is probably  lensed by foreground clusters and/or massive galaxies \citep[e.g.,][]{b110,b111}. 
 For instance, an ultra-bright mm/submm galaxy, MM J065837-5557.0, with a flux density of $\sim$ 20 mJy at 1100 $\mu$m \citep{b5} and $\sim$ 48 mJy at 870 $\mu$m \citep{b83} has been detected near the center of the Bullet cluster ($z=0.297$).
This source, whose brightness  appears to be highly boosted ($>$ 20--75) by gravitational lensing, is a luminous infrared galaxy (LIRG) behind the Bullet cluster at $z \sim 2.79$ \citep{b83,b104}. A similar but brighter source has also been reported toward the cluster MACS J2135-010217 ($z=0.325$). SMM J2135-0102 has an 870 $\mu$m flux density of 106 mJy, and is an ultra-luminous IR galaxy (ULIRG) at a spectroscopically confirmed redshift of 2.33 and with an amplification factor of $\sim$33 \citep{b74,b87}. 
Owing to the strong magnification of the gravitational lens, these lensed SMGs provide a unique opportunity to understand the physical properties of extreme star-formation in the early universe even with existing telescopes \citep{b74}.
Furthermore, SPT surveys have shown that  ultra-bright  submm/mm galaxies exist in some surface density; 
20 dust-dominated SMGs above 10 mJy at 1.4 mm were detected in 87 deg$^2$ area \citep{b71}.
The 14.4 deg$^2$ survey with Herschel, as a part of the H-ATLAS project,
uncovered 11 bright 500 $\mu$m sources ($>$ 100 mJy at 500 $\mu$m)
within the survey area, and 5 of 11 have been identified as lensed, dusty
starburst galaxies at $z = 1.6 - 3.0$ \citep{b126}.
Large mm/submm surveys enable us to find these new population.

Here, we report a serendipitous detection of an ultra-bright SMG in the Subaru/XMM-Newton Deep Field (SXDF),
termed as AzTEC-ASTE-SXDF1100.001 (hearafter referred to as SXDF1100.001) 
or Orochi\footnote{a Japanese word referring to an ancient Japanese legendary monster.}, 
during a course of wide and deep 1100 $\mu$m surveys using AzTEC mounted on ASTE.

This paper is organized as follows. The AzTEC on ASTE detection is reported in section 2,
and subsequent CARMA and SMA observations are described in section 3, along with the multi-wavelengths images from optical/infrared to radio.
Spectroscopic observations using Z-Spec on CSO  are discussed in section 4. 
The modeling of the SED and derived physical properties, including discussions on the source size/structure and implications for star formation properties of Orochi are given in section 5.

Throughout this paper, we adopt a cosmology with density parameters $\Omega_{\Lambda} = 0.7$ and $\Omega_{\rm M} = 0.3$ and the Hubble constant $H_{0} = 70$ km s$^{-1}$ Mpc$^{-1}$.

\section{AzTEC/ASTE 1100 $\mu$m observations and results}
\subsection{AzTEC/ASTE observations}
We conducted 1100 $\mu$m imaging observations of a cluster of Ly$\alpha$ Emitters (LAEs) at $z \sim 5.7$ \citep[Clump B;][]{b1} in SXDF 
using the AzTEC camera \citep{b4} mounted on ASTE \citep[][2008]{b35}, 
from November 26 to December 21, 2008. The observations were carried out remotely from the ASTE operation rooms  through the network observation system N-COSMOS3 developed by the National Astronomical Observatory of Japan (NAOJ) \citep{b3}. The full width at half maximum (FWHM) of the AzTEC beam on ASTE is $30^{\prime \prime}$ at 1100 $\mu$m, and the field of view of the array is roughly circular with a diameter of $8^{\prime}$.
During the ASTE 2008 observation run, 117 of  144 AzTEC detectors were operational.

We imaged a $\sim$6.6$^{\prime}$ diameter circular  field of Clump B, centered at R.A. (J2000) = $02^h 18^m 19.60^s$, Dec. (J2000) = $-5^{\circ} 32^{\prime} 52^{\prime\prime}.0$. 
We used a Lissajous scan pattern \citep{b5} in order to maximize the observation efficiency.
We selected a maximum velocity of $300^{\prime \prime} \mathrm{s^{-1}}$ in order to mitigate low-frequency atmospheric fluctuations.
We obtained a total of 39 individual observations for Clump B,  taking $\sim$40 min for  each observation. The atmospheric zenith opacity at 220 GHz was $\tau_{220 \mathrm{GHz}}=0.017$--$0.100$ as monitored with a radiometer at the ASTE telescope site.

Uranus or Neptune were observed at least once a night in order to measure each detector's point spread function (PSF) and relative position and to determine the flux conversion factor  for absolute calibration \citep{b4}. Pointing observations with the quasar J0132-169 were performed every 2h across observations for Clump B; the resultant pointing accuracy is better than $3^{\prime \prime}$ \citep{b5}. A pointing model is devised by interpolating these pointing data temporally and is applied to the astrometry for correction of pointings. Observational information is summarized in Table \ref{aztecobs}.

\subsection{AzTEC Data Reduction}
The data were reduced using the AzTEC data reduction pipeline written in the Interactive Data Language (IDL) in a manner similar to that in \citet{b6}. 
Here, we provide a brief summary of the process and point out the difference.
The data were divided into 15 s intervals of time-series data.
Spikes were then removed from the time-series data. 
A principal component analysis (PCA) method was used to subtract the sky emission, and the effect of PCA cleaning on the point source response is traced by reducing synthetic,
noiseless time-series data with a simulated point source using the same approach that employed for the actual data.
This `point source kernel' is indicative of the effect of PCA on the point source profile and flux attenuation of real point sources in the map 
and we corrected flux densities of AzTEC sources based on this kernel \citep{b128}.
The cleaned time-series data were projected into a map space using $3^{\prime \prime} \times 3^{\prime \prime}$ pixels, and the 39 individual observations were co-added into a single map by weighted averaging.
We also created 100 synthesised noise realizations by randomly multiplying each 15 s time-series interval by $\pm$1 (similar to the `scan-by-scan' jackknifing technique used in \citet{b6}).
These noise realizations are free of astronomical signals, including the signals from confused sources in the map.
The co-added map and the 100 noise realizations were  optimally filtered for the detection of point sources.

\begin{center}
\begin{table}
\caption{AzTEC on ASTE observations.}
\begin{tabular}{c c}
\hline
\hline
Parameters & Values \\ \hline
Observation date & Nov. 26 - Dec. 21, 2008 \\
Wavelength/Frequency & 1100 $\mu$m/270 GHz \\
Bandwidth       & 50  GHz \\
Number of detectors & 144 (total), 117 (operational) \\
Beam size (FWHM) & 30$^{\prime\prime}$ \\
Field center (J2000) & R.A. = $02^h 18^m 19.59^s$  \\
                     &  Dec. = $-05^{\circ} 32^{\prime} 52^{\prime\prime}.00$ \\
Field area          & 136 $\mathrm{arcmin^2}$ (50 \%-coverage region)\\
Map noise level     & 0.6--1.0 mJy \\
Pointing source & J0132--169 \\
Flux calibrator source & Uranus, Neptune \\
Opacity at 220 GHz & 0.017--0.100 \\
\hline
\end{tabular}
\label{aztecobs}
\end{table}
\end{center}

\subsection{AzTEC 1100 $\mu$m results}
The achieved average noise level of the resulting map  is 0.60--1.00 mJy over a 136 arcmin$^2$ area.
We found an ultra bright source, Orochi, at
$\alpha (J2000) = 02^{h} 18^{m} 30.68^{s}$ and $\delta (J2000) = -05^{\circ} 31^{\prime} 31^{\prime \prime}.37$. The flux density at 1100 $\mu$m is   37.28 $\pm$ 0.65 mJy. 
The 1100 $\mu$m image of Orochi is shown in Fig. 1. The source size is consistent with the point source kernel or PSF of the AzTEC/ASTE observations.
Hence, the point source kernel is understood as the AzTEC/ASTE beam.
The estimated position  errors from AzTEC signal-to-noise ratio and the beam size are $0^{\prime\prime}.9$ (1 $\sigma$) and 
$1^{\prime\prime}.4$ (2 $\sigma$).

\begin{figure}
\includegraphics[width=90mm]{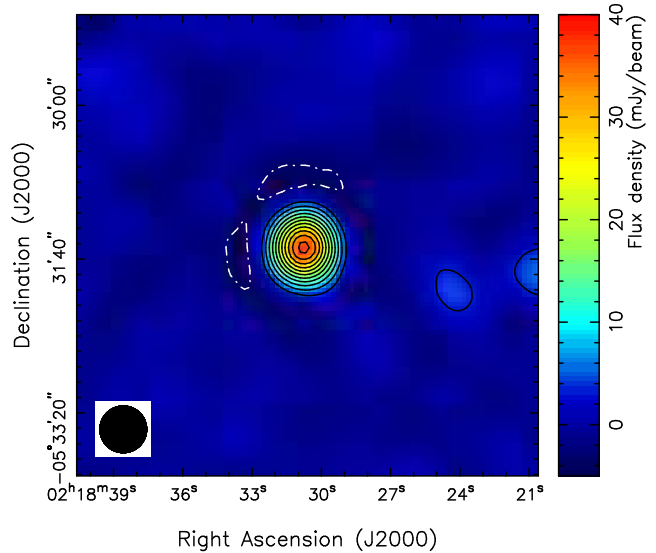}
\caption{The AzTEC/ASTE 1100 $\mu$m image of Orochi. The contour levels are 5, 10, 15, 20, 25, 30, 35, 40, 45, 50 and 55 $\sigma$, where 1 $\sigma$ = 0.65 mJy beam$^{-1}$. Negatives are shown in white contours.} 
\label{azmap}
\end{figure}

\section{Interferometric identification and multi wavelength properties}
We performed interferometric imaging with CARMA and SMA of Orochi to constrain its position.
With the refined position estimate, we searched the literature and the archives for multi-wavelength data.
\subsection{CARMA 1300 $\mu$m observations and results }

\subsubsection{CARMA 1300 $\mu$m Observations \& Reduction}
We conducted follow-up observations of Orochi on August 16, 2009, using the D configuration with 15 antennas of the CARMA. The phase centre was  $\alpha (\mathrm{J2000}) = 02^{h} 18^{m} 21^{s}$ and $\delta (\mathrm{J2000}) = -05^{\circ} 31^{\prime} 31^{\prime \prime}$. The projected baseline lengths ranged from 10 m to 108 m. The CARMA correlator was configured to cover a  1.5 GHz width in each sideband, yielding a total bandwidth of 3 GHz for continuum observations after adding the two sidebands.
The centre frequency of the receivers was tuned to 230 GHz.
We observed a bright QSO, J0108+015 ($18.8^{\circ}$ away from Orochi) as a visibility calibrator and Uranus as a flux calibrator.
To obtain an empirical upper limit on the systematic position error induced by baseline errors, we observed a radio galaxy, J0241-082 ($25^{\circ}$ away from the visibility calibrator; the distance is approximately 1.33 times that between  Orochi and the visibility calibrator) during  a track. We observed a bright QSO, 3C84, as a bandpass calibrator.
The raw CARMA data were calibrated and imaged with natural weight using MIRIAD \citep{b23}.
We found that the quality of the data from the upper side band (USB) was significantly poorer than that of  the lower side band (LSB) data. Therefore, we used only the LSB data for pinpointing the position of Orochi and subsequent analysis.
Observational information is summarized in Table 2.

\begin{center}
\begin{table}
\caption{CARMA observations.}
\begin{tabular}{c c}
\hline
\hline
Parameters &  Values \\ \hline
Observation date & Aug. 16, 2009 \\
Wavelength/Frequency & 1300 $\mu$m/232 GHz \\
Bandwidth           & 1.5 GHz \\
Phase center (J2000) & R.A. = $02^{h} 18^{m} 21^{s}$ \\
                     &  Dec. = $-05^{\circ} 31^{\prime} 31^{\prime \prime}$\\
Phase calibrator     & J0108+015 \\
Flux calibrator      & Uranus \\
Array configuration  & D configuration \\
projected baseline & 10--108 m \\
Primary beam & 35$''$.4 (FWHM) \\
Synthesized Beam size & $3^{\prime \prime}.2 \times 2^{\prime \prime}.1$ (P.A.$ =-16.4^{\circ}$) \\
Map noise level     & 1.3 mJy \\
Opacity at 230 GHz & 0.1--0.3 \\
\hline
\end{tabular}
\label{carmaobs}
\end{table}
\end{center}

\subsubsection{CARMA 1300 $\mu$m results}

\begin{figure}
\includegraphics[width=90mm]{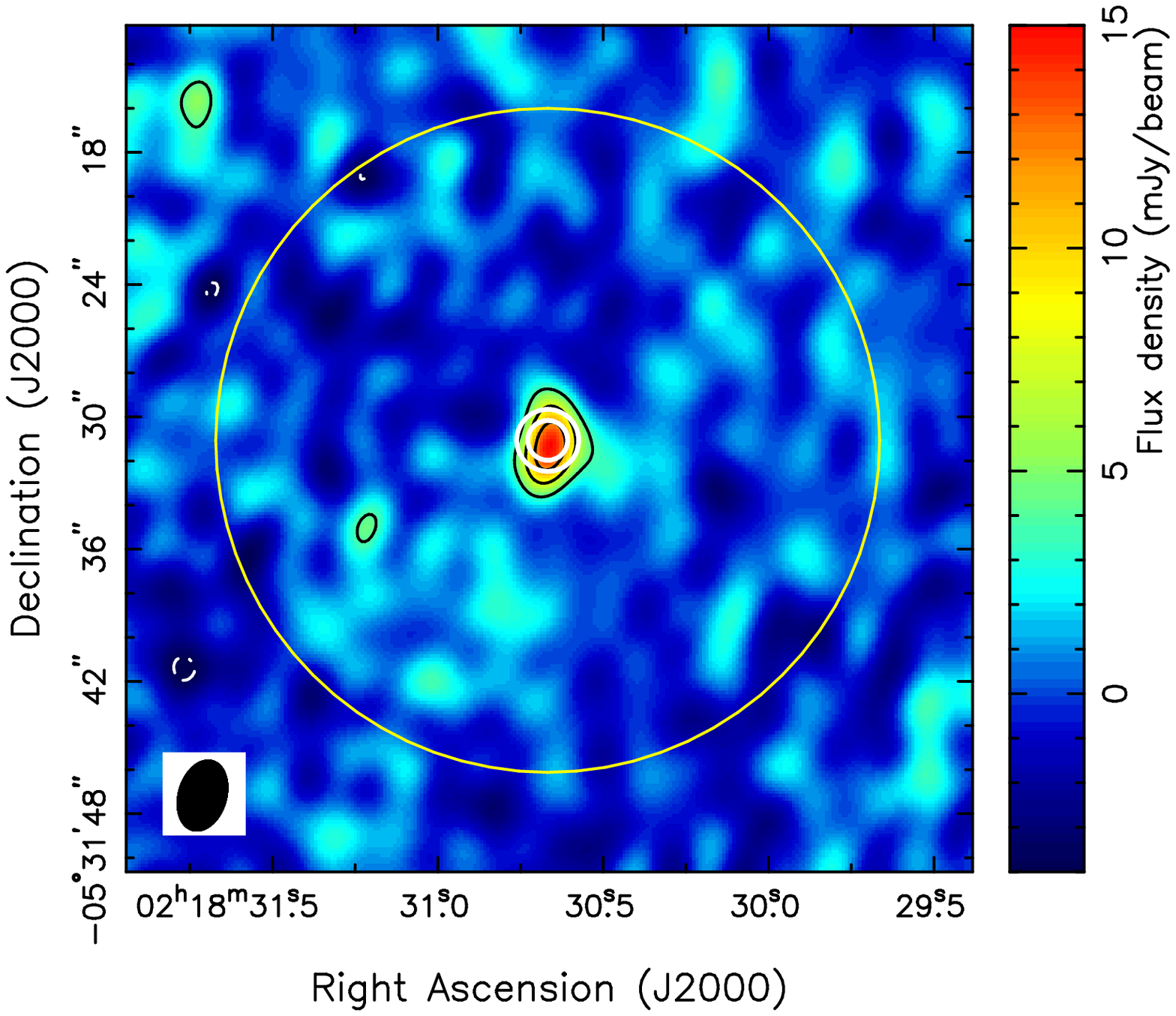}
\caption{The CARMA 1300 $\mu$m image of Orochi. The contour levels are 3, 6, and 9 $\sigma$ (1 $\sigma$ = 1.3 mJy). Negatives are shown in white contours. The synthesized beam is shown on the lower left.
Note that no correction of the primary beam attenuation was made in this map.
The larege solid yellow circle shows the beam size of AzTEC/ASTE ($30^{\prime\prime}$). The small solid white circles show the estimated position errors; the inner one shows 1 $\sigma$ error ($0^{\prime\prime}.9$) and the outer one shows 2 $\sigma$ error ($1^{\prime\prime}.4$).} 
\label{carmap}
\end{figure}

We find a source with 11 $\sigma$ significance at 
$\alpha (J2000) = 02^{h} 18^{m} 30.67^{s}$ and $\delta (J2000) = -05^{\circ} 31^{\prime} 31^{\prime \prime}.42$ (Fig. \ref{carmap}).
The CARMA 1300 $\mu$m source position  coincides well with that of  the centroid of the AzTEC/ASTE 1100 $\mu$m source.
The derived source position is shown in Table \ref{carmaresult}.

To evaluate the reliability of the astrometry in our CARMA observations, we estimate the statistical errors due to the noise obtained in fitting a point source to the calibrated visibilities and systematic errors due to uncertainties
 in interferometer baselines.
Details of these analysis followed  \citet{b27}.
The derived statistical errors are shown in Table \ref{carmaresult}.
The systematic astrometry error caused by baseline length uncertainties is estimated as $\sim$0$^{\prime\prime}.5$ based on the 
CARMA image of a known radio source J0241-082.
The estimated statistical positional errors for RA and Dec are 0$^{\prime\prime}$.09 and $0^{\prime\prime}$.12.

\begin{table*}
\caption{Peak positions and their errors for VLA 20 cm, CARMA 1300 $\mu$m, AzTEC 1100 $\mu$m, SMA 880 $\mu$m, and Subaru $z^{\prime}$-band images of Orochi. The measured offsets from the $z'$-band peak position are also listed.}
\begin{tabular}{c c c c c c c c c}
\hline
                    &                           &                                                     & $\sigma(\alpha)$&  $\sigma(\delta)$ & $\sigma^{\prime}(\alpha)$$^a$&  $\sigma^{\prime}(\delta)$$^a$ &$z^{\prime}$-band offset $\sigma(\alpha)$  & $z^{\prime}$-band offset $\sigma(\delta)$ \\ 
image               & $\alpha$(J2000)          & $\delta$(J2000)                                    &  (arcsec)         &   (arcsec)         &  (arcsec)                    &             (arcsec)         & (arcsec)                                  & (arcsec)                                       \\ \hline
 VLA                 & $02^{h} 18^{m} 30.67^{s}$ & $-05^{\circ} 31^{\prime} 31^{\prime \prime}.48$  &   0.07          &     0.09           &      0.26                        &    0.23            &  $-$0.18        &   0.32            \\           
CARMA               & $02^{h} 18^{m} 30.67^{s}$ & $-05^{\circ} 31^{\prime} 31^{\prime \prime}.28$  &  0.09            &   0.12             &      0.27                        &    0.24            &  $-$0.18                                       &   0.52                                         \\
AzTEC                 & $02^{h} 18^{m} 30.67^{s}$ & $-05^{\circ} 31^{\prime} 30^{\prime \prime}.97$  &  0.9             &   0.9            &       0.9                       &    0.9             &  $-$0.18                    &   0.83                        \\     
 SMA               & $02^{h} 18^{m} 30.68^{s}$ & $-05^{\circ} 31^{\prime} 31^{\prime \prime}.68$  &  0.04            &   0.06             &        0.25                       &    0.22             &  0.00                                       &   0.12                                         \\
 $z^{\prime}$-band & $02^{h} 18^{m} 30.68^{s}$ & $-05^{\circ} 31^{\prime} 31^{\prime \prime}.80$ &   0.25          &  0.21              &          -                     &        -          &                -                        &                             -                \\ \hline
\end{tabular}
\\$^a$ denotes the combined ($z^{\prime}$ plus radio or submm/mm) 1 $\sigma$ uncertainty in position.
\label{carmaresult}
\end{table*}

\subsection{SMA 880 $\mu$m observations and results}
\subsubsection{SMA 880 $\mu$m Observations \& Reduction}
We conducted follow-up observations of Orochi on December 10, 2009, using the compact configuration with 8 antennas of the SMA. The phase center was $\alpha (\mathrm{J2000}) = 02^{h} 18^{m} 22^{s}$ and $\delta (\mathrm{J2000}) = -05^{\circ} 31^{\prime} 35^{\prime \prime}$.
The projected baseline lengths ranged from 7 m to 70 m. The SMA correlator was equipped with 4.0 GHz in each sideband, providing a total of 8 GHz bandwidth for continuum observations.
The centre frequency of the receivers was tuned to 340 GHz.
We observed a bright QSO, J0132-169 ($16.0^{\circ}$ away from Orochi) as a visibility calibrator and Uranus  as a  flux calibrator.
To obtain an empirical upper limit on the systematic position error induced by baseline errors, we observed a radio galaxy, J0238-166 ($22.7^{\circ}$ away from the visibility calibrator; the distance is approximately 1.4 times that between Orochi and the visibility calibrator) during  a track.
We observed a bright QSO, 3C454.3, and  Callisto as bandpass calibrators  during the track.

The raw SMA data were calibrated using the MIR package \citep{b22}. 
Imaging was carried out in MIRIAD with natural weighting. 
Observational information is summarized in Table 4.

\begin{center}
\begin{table}
\caption{SMA observations.}
\begin{tabular}{c c}
\hline
\hline
Parameters &  Values \\ \hline
Observation date & Dec. 10, 2009 \\
Wavelength/Frequency & 880 $\mu$m/ 340 GHz \\
Bandwidth           & 8 GHz \\
Phase center (J2000) & R.A. = $02^{h} 18^{m} 22^{s}$ \\
                     &  Dec. = $-05^{\circ} 31^{\prime} 35^{\prime \prime}$\\
Phase calibrator     & J0132-169 \\
Flux calibrator      & Uranus \\
Array configuration  & compact configuration \\
projected baseline & 7--70 m \\
Primary beam & 32$''$.4 (FWHM) \\
Synthesized Beam size & $3^{\prime \prime}.0 \times 1^{\prime \prime}.9$ (P.A.$ =8.1^{\circ}$) \\
Map noise level     & 2.2 mJy \\
Opacity at 225 GHz & 0.1--0.3 \\
\hline
\end{tabular}
\label{smaobs}
\end{table}
\end{center}

\subsubsection{SMA 880 $\mu$m results}

\begin{figure}
\includegraphics[width=90mm]{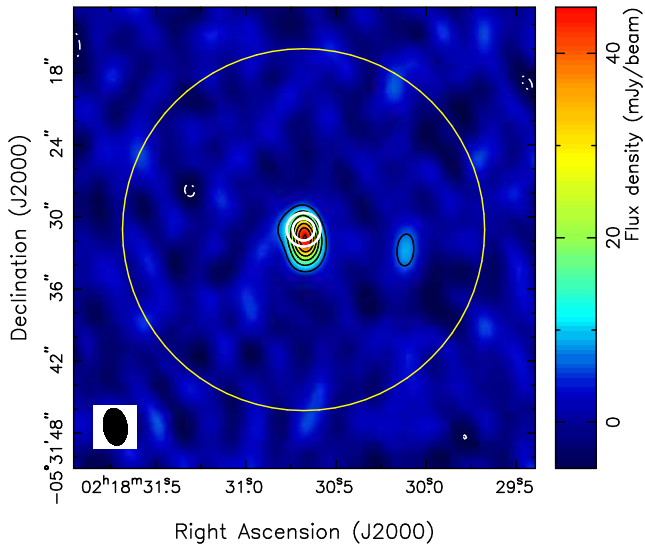}
\caption{The SMA 880 $\mu$m image of Orochi. The contour levels are -3, 3, 6, 9, 12, 15, 18, and 21 $\sigma$ (1 $\sigma$ = 2.2 mJy). Negatives are shown in white contours. The synthesized beam is shown on the lower left. Note that no correction of the primary beam attenuation was made in this map. The large solid yellow circle shows the beam size of AzTEC/ASTE ($30^{\prime\prime}$). The small solid white circles show the estimated position errors; the inner one shows 1 $\sigma$ error ($0^{\prime\prime}.9$) and the outer one shows 2 $\sigma$ error ($1^{\prime\prime}.4$).} 
\label{smamap}
\end{figure}

We find a source with 21 $\sigma$ significance at the position of $\alpha (J2000) = 02^{h} 18^{m} 30.67^{s}$ and $\delta (J2000) = -05^{\circ} 31^{\prime} 31^{\prime \prime}.68$ (Fig. \ref{smamap}).
The SMA 880 $\mu$m source position coincides with that of the centroid of the AzTEC/ASTE 1100 $\mu$m source and CARMA 1300 $\mu$m source.
The derived source positions of AzTEC/ASTE, CARMA, and SMA are shown in Table \ref{carmaresult} along with the VLA source position.

The systematic astrometry error of the SMA source caused by baseline length uncertainties was estimated as $\sim$0$^{\prime\prime}.1$ based on the 
SMA image of  J0238-166.
The estimated statistical positional errors for RA and Dec are 0$^{\prime\prime}$.04 and 0$^{\prime\prime}$.06.

\subsection{Multi-wavelength Data}
\begin{figure*}
\includegraphics[width=160mm]{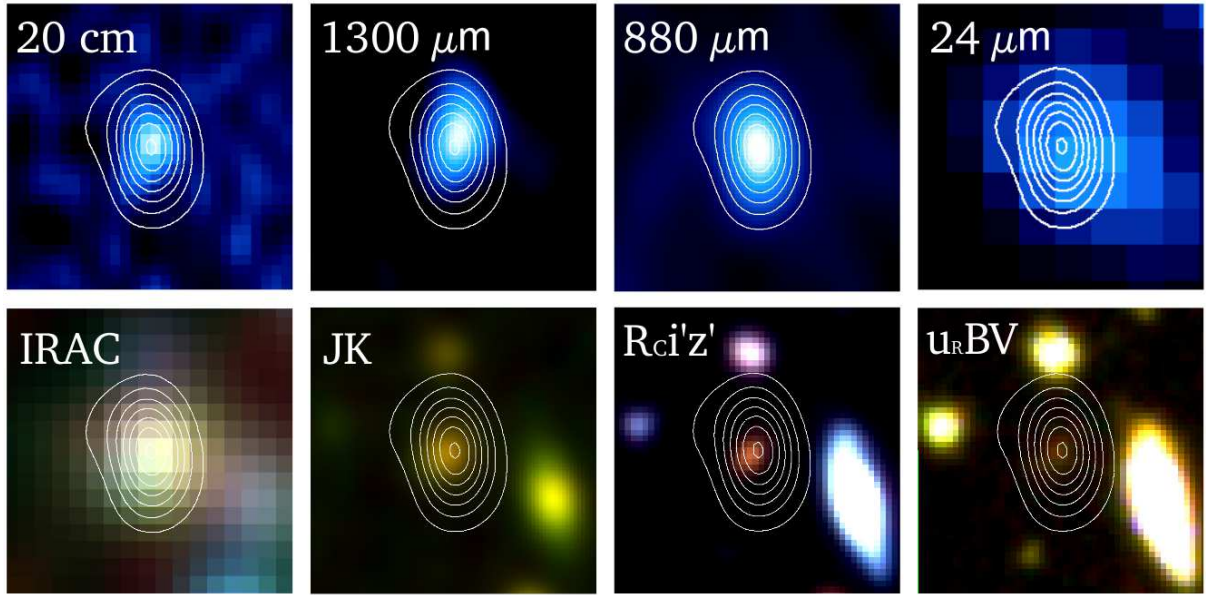}
\caption{Multi-wavelength images of Orochi with SMA contours (3, 6, 9, 12, 15, 18, and 21 $\sigma$). The size of each image is $10^{\prime \prime} \times 10^{\prime \prime}$; north is towards the top of the image and east is towards  the left side of the image. From left to right and top to down: VLA 20 cm; CARMA 1300 $\mu$m; SMA880 $\mu$m; MIPS 24 $\mu$m; rgb image of IRAC/ch1 (blue), ch2 (green), ch3 and ch4 (red); rg image of WFCAM/$J$ (green) and $K$ (red); rgb image of SuprimeCam/$R$, $i^{\prime}$ and $z^{\prime}$; rgb image of SuprimeCam/$u_R$, $B$ and $V$.} 
\label{stamp}
\end{figure*}

We find a multi-wavelength counterpart at the CARMA/SMA position using optical, near-infrared (NIR), mid-infrared (MIR), and radio wavelengths from archival data.
The derived flux and corresponding images are displayed in Table 5 and Fig. 4, respectively. Here, we briefly describe the archival data.
Optical data in 5 broad-bands, $B$, $V$, $Rc$, $z'$, and $i'$-band, were provided by the SXDS project data release 1 \citep{b2}.
A $u_R$-band image was obtained from the SMOKA archive in Japan.
$K$-band and $J$-band images were provided by UKIDSS/UDS surveys data release 1 \citep{b37}.
 Infrared data in all four Spitzer IRAC bands (3.6, 4.5, 5.8, and 8 $\mu$m) are available from the SWIRE survey, and the MIPS band (24 $\mu$m) is available from the SpUDS survey. These data were obtained from the Spitzer archive.  
Orochi was detected in a 20 cm VLA image and catalogued as VLA J021830-05315 \citep{b38}. An updated VLA image, taken in the A-configuration (Arumugam et al., in prep.), has been incorporated into the multi-wavelengths dataset.
Unfortunately, the counterpart of Orochi is not isolated on the 24 $\mu$m image  
. Other infrared bright source is located in south west of Orochi and it looks like an extended source in the optical wavelength images (Fig. \ref{stamp}). 
Therefore, we regard the aperture photometry data in the SWIRE catalogue as an upper limit.
The photometry data at 1000, 1100, 1200, 1300, 1400, and 1500 $\mu$m are obtained from Z-Spec observations. See section 4 for details.

\begin{table*}
\begin{minipage}{125mm} 
\caption{Flux densities of Orochi.}
\setlength{\tabcolsep}{0.8mm}
\begin{tabular}{l c c c}
\hline
Wavelength &Camera/Telescope & Magnitude/Flux & Reference \\ \hline
$4.5-10 \ \mathrm{keV}$ &EPIC/XMM-Newton & $<$ 5 $\times 10^{-15}$($\mathrm{erg/cm^2/s}$)$^a$ &\citet{b10}\\
$0.5-4.5 \ \mathrm{keV}$ &EPIC/XMM-Newton & $<$ 8 $\times 10^{-16}$($\mathrm{erg/cm^2/s}$)$^a$ &\citet{b10}\\
$0.380 \ \mathrm{\mu m}$ ($u_R$-band) &SuprimeCam/Subaru& $<$ 26.76 ($m_{AB}$)$^b$ &-\\
$0.445 \ \mathrm{\mu m}$ ($B$-band) &SuprimeCam/Subaru& 25.86 $\pm$  0.05 ($m_{AB}$) &\citet{b2}\\
$0.551 \ \mathrm{\mu m}$ ($V$-band) &SuprimeCam/Subaru& 25.51 $\pm$  0.05 ($m_{AB}$)& \citet{b2}\\
$0.659 \ \mathrm{\mu m}$ ($Rc$-band) &SuprimeCam/Subaru& 24.80 $\pm$ 0.05 ($m_{AB}$)& \citet{b2}\\
$0.771 \ \mathrm{\mu m}$ ($i^{\prime}$-band) &SuprimeCam/Subaru& 23.68 $\pm$ 0.05 ($m_{AB}$) &\citet{b2}\\
$0.922 \ \mathrm{\mu m}$ ($z^{\prime}$-band) &SuprimeCam/Subaru & 22.88 $\pm$ 0.05 ($m_{AB}$) &\citet{b2}\\
$1.215 \ \mathrm{\mu m}$ ($J$-band) & WFCAM/UKIRT & 0.0096 $\pm$ 0.00066 (mJy)&\citet{b37}\\ 
$2.179 \ \mathrm{\mu m}$ ($K$-band) & WFCAM/UKIRT & 0.0224 $\pm$ 0.0017 (mJy)&\citet{b37}\\ 
$3.6 \ \mathrm{\mu m} $ &IRAC/\textit{Spitzer} & 0.046 $\pm$ 0.0018 (mJy) &-\\
$4.5 \ \mathrm{\mu m} $ &IRAC/\textit{Spitzer} & 0.057 $\pm$ 0.0017 (mJy) &-\\
$5.8 \ \mathrm{\mu m} $& IRAC/\textit{Spitzer} &0.075 $\pm$ 0.0088 (mJy) &-\\
$8 \ \mathrm{\mu m}   $  & IRAC/\textit{Spitzer} & 0.110 $\pm$ 0.013 (mJy) &-\\
$24 \ \mathrm{\mu m} $ & MIPS/\textit{Spitzer} &$<$ 0.384 (mJy)& -\\
$70 \ \mathrm{\mu m} $ & MIPS/\textit{Spitzer} &$<$ 12 (mJy)$^c$& -\\
$160 \ \mathrm{\mu m} $ & MIPS/\textit{Spitzer} &$<$ 72 (mJy)$^c$& -\\
$880 \ \mathrm{\mu m} $ & -/SMA & 90.7 $\pm$ 2.2 $\pm$ 20.7$^d$ (mJy)& This work\\
$1000 \ \mathrm{\mu m}$ & Z-SPEC/CSO & 51.9 $\pm$ 1.2 (mJy) & This work  \\
$1100 \ \mathrm{\mu m}$ & AzTEC/ASTE  & 37.3  $\pm$ 0.7 (mJy) & This work\\
$1100 \ \mathrm{\mu m}$ & Z-SPEC/CSO & 38.4 $\pm$ 0.9 (mJy) & This work  \\
$1200 \ \mathrm{\mu m}$ & Z-SPEC/CSO & 31.3 $\pm$ 0.8 (mJy) & This work  \\
$1300 \ \mathrm{\mu m}$ & -/CARMA  & 24.6 $\pm$ 1.3 $\pm$ 5.3$^d$  (mJy)& This work \\ 
$1300 \ \mathrm{\mu m}$ & Z-SPEC/CSO & 23.9 $\pm$ 1.0 (mJy) & This work  \\
$1400 \ \mathrm{\mu m}$ & Z-SPEC/CSO & 19.5 $\pm$ 0.9 (mJy) & This work  \\
$1500 \ \mathrm{\mu m}$ & Z-SPEC/CSO & 13.9 $\pm$ 1.2 (mJy) & This work  \\ 
$20 \ \mathrm{cm} $ &-/VLA & 0.238 $\pm$  0.038 (mJy) &  Arumugam et al., in prep   \\ 
$50 \ \mathrm{cm} $ &-/GMRT & 0.315 $\pm$ 0.120 (mJy) & \citet{b102}  \\  \hline
\end{tabular}
\\ $^a$ indicates the detection limit in Ueda et al. 2008 and corresponds to a confidence level of 99.91\%.
\\ $^b$ indicates the detection limit with 2 $\sigma$ and aperture size of $2^{\prime\prime}$ (diameter).
\\ $^c$ indicates the upper limit with $3 \sigma$. 
\\ $^{d}$ indicates the uncertainty of uv model fitting for the extended structure. See section 5.1 for details.
\label{flux}
\end{minipage} 
\end{table*}

\section{Z-Spec 1500 $\mu$m to 1000 $\mu$m ultra-wideband spectroscopy}

\subsection{Observations and data reduction}

In order to determine the redshift of Orochi, we searched for redshifted molecular/atomic lines using Z-Spec, a single-beam grating spectrometer which disperses the 190--308 GHz (or 1570--970 $\mu$m) band across a linear array of 160 bolometers \citep[][and references therein]{b94}.  
The resolving power (R) runs from 250 to 300, from the highest end to the lowest end of the band. This gives a frequency resolution of $\sim$ 700 MHz or a velocity resolution of $\sim$ 880 km s$^{-1}$ at the centre of the band ($\sim$ 240 GHz).

Orochi was observed for six nights during late November to early December of 2009 using Z-Spec mounted on the Caltech Submillimeter Observatory (CSO) 10 m telescope. A traditional chop-and-nod mode, with the secondary chopping at 1.6 Hz and a nod period of 20 s, was adopted. The total integration time was 16 h, excluding bad scans (showing unreasonable baseline shape and/or offset). The absolute intensity scale was calibrated from observations of Mars, with a channel-to-channel correction based on quasar spectra measured during the observation run. The overall calibration uncertainties are estimated to be less than 10\%, except at the lowest frequencies which are degraded by the wing of the 186 GHz atmospheric water line. The Z-Spec observations are summarized in Table 6.

\begin{center}
\begin{table}
\caption{Z-Spec observations.}
\begin{tabular}{c c}
\hline
\hline
Observation date & Values \\ \hline
Observation date & Nov.--Dec., 2009 \\
Wavelength/Frequency &  970--1570 $\mu$m/ 190--308 GHz \\
Spectral resolution  &  $\sim$ 700 MHz at 240 GHz \\
Flux calibrator      & Mars   \\
Averaged noise level     & $\sim$ 2.7 mJy \\
Opacity at 225 GHz & 0.03--0.17 \\
\hline
\end{tabular}
\label{zspecobs}
\end{table}
\end{center}

\subsection{Z-Spec result}
\begin{figure*}
\includegraphics[width=160mm]{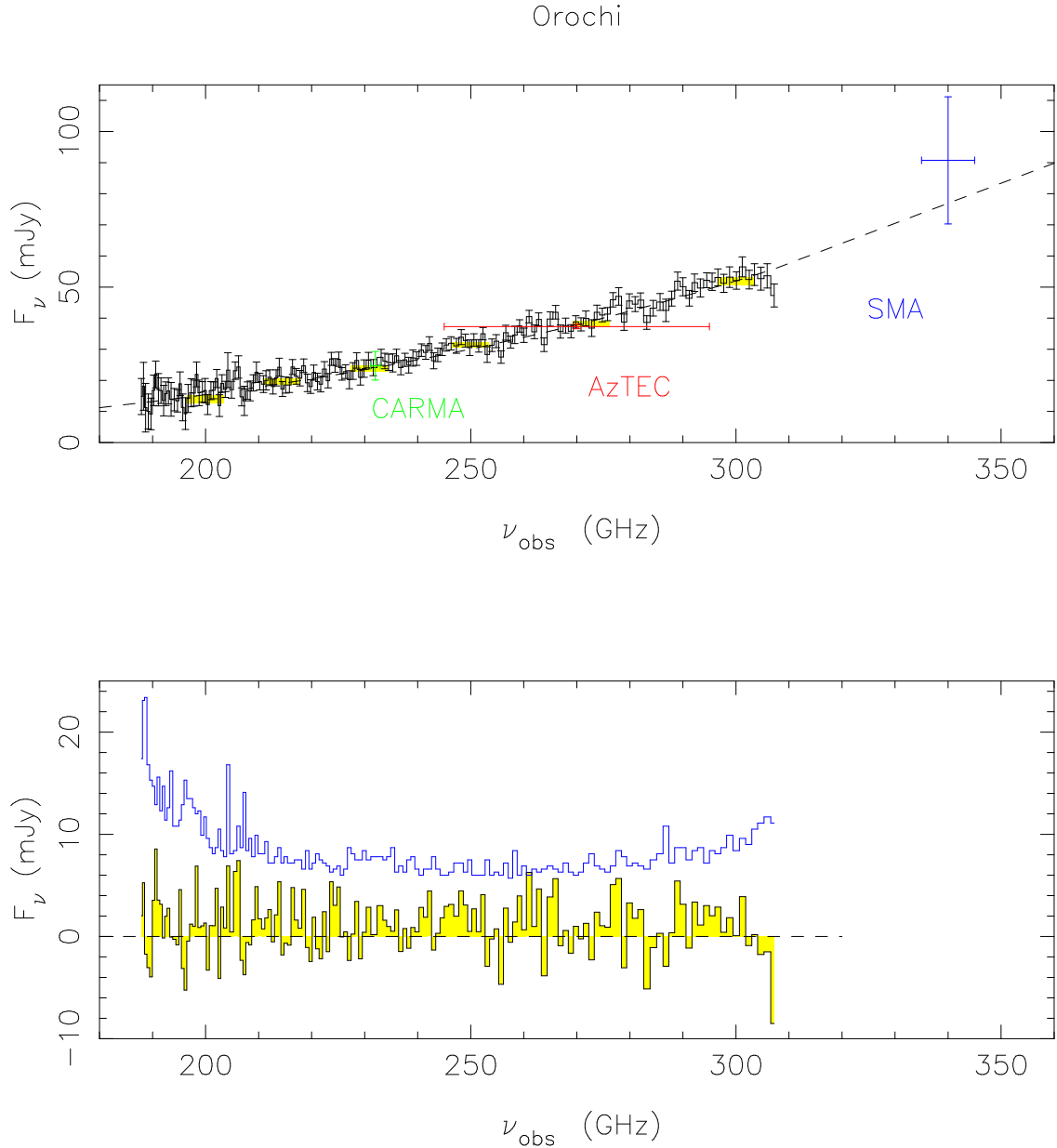}
\caption{The  Z-Spec mm-wavelength spectrum of Orochi. The top panel shows observed flux density by Z-Spec. The dotted line is the fitted continuum, S $\propto \nu^{3.00 \pm 0.11}$. The blue, red and green data points show the flux densities and error bars (without absolute calibration error) measured with the  SMA, AzTEC/ASTE, and CARMA, respectively. The horizontal bars give the bandwidth. The yellow box shows areas used for calculating the continuum flux densities in Table 5. 
The bottom panel shows the Z-Spec spectrum after subtracting continuum flux.
The yellow histogram shows the residual spectrum of Orochi and the blue histogram shows the 3 $\sigma$ errors.
} 
\label{zspec}
\end{figure*}

The derived Z-Spec spectrum of Orochi is displayed in Fig. \ref{zspec}. Continuum emission is detected across the whole band, and the S/N ratio exceeds 10 above $\sim$220 GHz. The averaged noise level of the spectrum is 2.7 mJy. We found a clear slope of continuum emission, which is attributed to the thermal dust emission. The best-fit continuum spectrum across the Z-Spec band is
\begin{equation}
F_{\nu} = \left( 30.1 \pm 0.5 \right) \left( \frac{\nu}{250 \mbox{\ \ GHz} }\right)^{3.00 \pm 0.11} \mbox{mJy}.
\end{equation}
Considering absolute flux calibration errors of 10--20\% in the mm wavelength band, the measured flux densities with SMA and CARMA coincide well with the Z-Spec continuum flux. 
We  derived the average flux densities  at 1000, 1100, 1200, 1300, 1400, and 1500 $\mu$m by binning  the Z-Spec spectrum into bands of width  $\sim$7 GHz. The results are listed in Table \ref{flux}.


We see no significant emission/absorption line feature across the band, and therefore we were unable to determine the redshift of Orochi from these Z-Spec data. For a more quantitative analysis, we use the redshift-finding algorithm described in Lupu et al. (in prepration.), which computes the average S/N ratio from all CO, [CI], [CII], and [NII] transitions falling into the Z-Spec bandpass as a function of redshift. The resulting redshift probability function is very noisy, and the average S/N ratio at all redshifts is $<$ 3 sigma, demonstrating the lack of significant line emission in these data.

The upper limits on the expected CO emission lines and their implications are discussed in section 5. 

\section{Discussion}
In this section, we discuss 
the properties of Orochi, an ultra-bright SMG identification in AzTEC/ASTE surveys and confirmed by subsequent CARMA/SMA observations.  
First, the source structure is; the discovery of the spatially extended submm/mm bright component accompanied with an unresolved compact bright source is discussed. Then, detailed  modeling of SEDs as well as the derivations of physical quantities are presented. 
We suggest that an ultra-bright SMG at a redshift of $\sim$3.4 has been discovered just behind a red foreground galaxy at $z$ $\sim$ 1.4 and that  Orochi is likely to be lensed by the foreground galaxy.

The deduced star formation properties and internal structures of Orochi are also discussed, along with their implications.

\subsection{Discovery of the extended  bright structure in Orochi}

\begin{figure*}
\includegraphics[width=190mm]{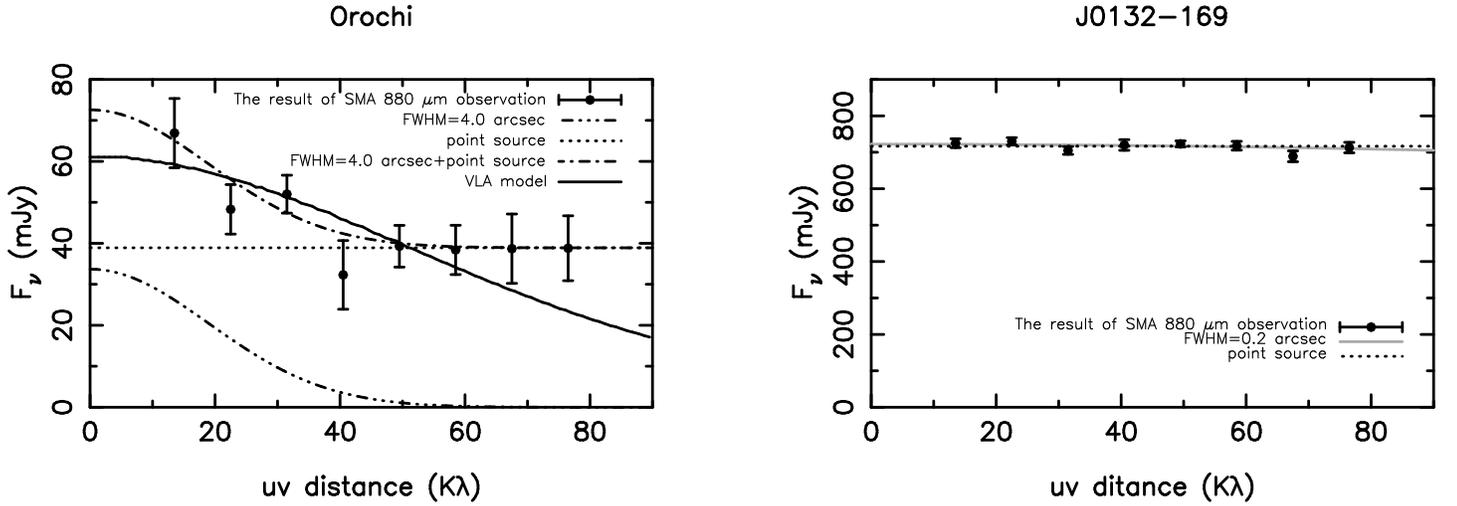}
\caption{880 $\mu$m visibility amplitude as a function of the baseline lengths (uv distances) toward Orochi (left) and the visibility calibrator (right),
along with the results of Gaussian  and point source fitting. These plots only show the real part of the visibilities, and the flux densities are not primary-beam corrected on these plots.
The visibility amplitude of the calibrator (right) is well modeled with an unresolved point source, whereas Orochi can be fitted with two components, i.e., a combination of a Gaussian shaped component with a FWHM of $\sim$4$^{\prime\prime}$ (dash-dot-dot-doted line) and an unresolved point source (dotted line). The resultant fitting is indicated by a dot-dashed line. 
The solid curve line shows the amplitude function derived from the VLA source size ($1^{\prime \prime}.84 \times 0^{\prime\prime}.95$ at PA 23.9$^{\circ}$), and its amplitude is scaled to be fitted to the SMA result.}
\label{smauv}
\end{figure*}

We find that Orochi is spatially resolved and that is has two components--- an extended structure and a compact unresolved one--- from the analysis of the visibility amplitude as a function of the projected baseline length in SMA data (Fig. \ref{smauv}). 
The plot of the phase calibrator is almost constant,
whereas the visibility amplitude of Orochi appears to  
reduce significantly for a longer uv-distance.
The visibility amplitude was fitted with a combination of a Gaussian (FWHM of $4^{\prime\prime}.0$)   and an unresolved  point source.
This fitting was carried out using the uvfit task in MIRIAD.
The estimated flux densities of the extended and compact structures are   33.7 $\pm$ 16.2 mJy and  38.9 $\pm$ 3.2 mJy, respectively;
in other words approximately a half of the total 880 $\mu$m flux ($\sim$73 mJy) is originated from the extended component.
Note that the  flux densities of the extended component and the unresolved component are 42.1 $\pm$ 20.3 mJy and 48.6 $\pm$ 4.0 mJy, respectively, after applying a correction of the primary beam response\footnote{In order to confirm the existence of the extended component, we carried out imaging   of the extended component from the residual data after subtracting the unresolved component in uv domain.
For this imaging, we used uv data at uv-distance $=10-30$ k$\lambda$ in order to achieve better signal-to-noise ratio.
The image achieve the noise level of 4.1 mJy and the synthesized beam size of 4$''$.8x4$''$.3 (P.A.=84$^{\circ}$).
We found a point source on the residual image and its flux density is 20.8 $\pm$ 4.1 mJy.
This flux density is consistent with the estimated flux density at uv-distance of 10-30 k$\lambda$ on the visibility amplitude plot, 18.9 $\pm$ 4.8 mJy (Fig. 6).
In applying primary beam correction for Orochi, the effect of the extended source size should be considered, because Orochi is located 9 arcsec away from the phase centre of SMA.
The primary beam response is 0.80 at the center of the source position, 0.87 at the inner edge of the extended components , and 0.73 at the outer edge of the extended components.
This means that the effect of primary beam correction for the extended components depends on its structure.
In order to evaluate this effect, we applied primary beam correction for the residual image. 
Considering the fact that the extended component is unresolved in the beam size of this map ($\sim 4''.7$), the corrected flux density on the residual image is equivalent to the corrected flux density of the extended component.
The corrected flux density of the residual image is 25 mJy, which is equivalent to applying the primary beam response of 0.8 , therefore we can apply the primary beam response of 0.8 to the extended component.}.

On the VLA image, Orochi is spatially resolved with a size of $1^{\prime \prime}.84 \times 0^{\prime\prime}.95$ at PA 23.9$^{\circ}$.
A single source model with this VLA source size can also fit the data, but the two-component model provides a better fit (Fig. \ref{smauv}).
The exact reason for the difference between the source size determined via our VLA and SMA data is not clear. The VLA image published by \citet{b38} was made with visibilities from the VLA's DnC and B configurations and yields a flux only 30 $\mu$Jy ($\sim$10 \%, and $<1\sigma$) below the deeper VLA image used here (Arumugam et al., in prep), which supplements the Simpson et al.\ data with visibilities from A configuration. It is unlikely, therefore, that any loss of significant extended structure is due to the $uv$ coverage. It is possible that the radio-to-submm flux ratio varies across the structure: a dust temperature gradient may produce such a difference, because a lower dust temperature results in a lower radio flux density, based on the radio-FIR correlation \citep[e.g.,][]{b16,b129} as shown in Fig.~12. It is also possible that faint, compact AGN-related emission may influence the radio morphology, leading to a smaller apparent size. It is clear that more sensitive submm and radio imaging is required to address the issue conclusively.

\begin{figure*}
\includegraphics[width=190mm]{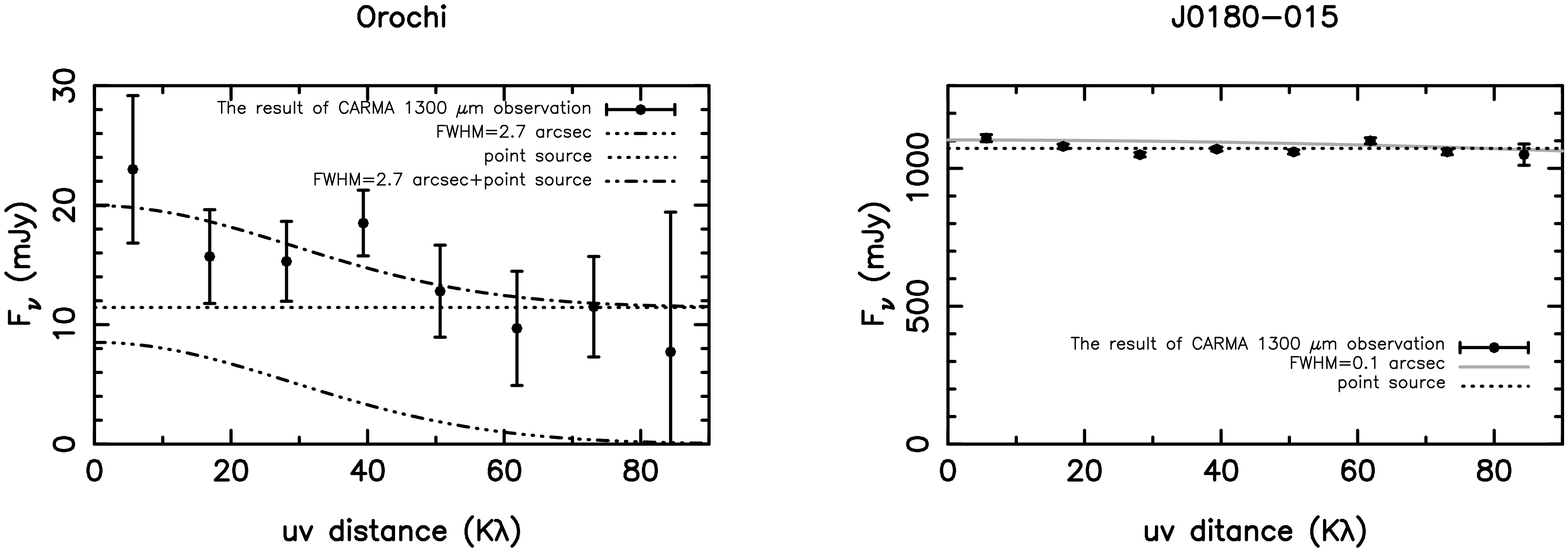}
\caption{1300 $\mu$m visibility amplitude as a function of the baseline lengths (uv distances) toward Orochi (left) and the visibility calibrator (right),
along with the results of Gaussian  and point source fitting. These plots only show the real part of the visibilities, and the flux densities are not primary-beam corrected on these plots.
The visibility amplitude of the calibrator (right) is well modeled with an unresolved point source, whereas Orochi can be fitted with two components, i.e., a combination of a Gaussian shaped component with a FWHM of $\sim$2$^{\prime\prime}$.7 (dash-dot-dot-doted line) and an unresolved point source (dotted line). The resultant fitting is indicated by a dot-dashed line. }
\label{uv}
\end{figure*}

This two-component feature can also be observed in the CARMA 1300 $\mu$m image (Fig. \ref{uv}), although the signal-to-noise ratio of the CARMA data is worse than that of the SMA data.
We conducted two-component fittings of the CARMA data as we did for the SMA data.
The resultant source size of the extended component on the CARMA data is $\sim$2$^{\prime\prime}.7$ (FWHM).
The estimated flux densities of the extended and compact structures are   8.5 $\pm$ 3.5 mJy and  11.4 $\pm$ 2.5 mJy, respectively, and the total flux density is $\sim$20 mJy at 1300 $\mu$m  (Table \ref{size}).
We also applied primary beam correction for CARMA image using a primary beam response of 0.81, as we did for SMA image, because Orochi is located 10 arcsec away from phase center on CARMA image. The corrected flux densities of the extended component and the unresolved component are 10.5 $\pm$ 4.3 mJy and 14.1 $\pm$ 3.1 mJy, respectively.
The estimated source size of the extended component by SMA and CARMA is also listed in Table 7.
We suggest that the SMA and CARMA source sizes are consistent if we consider the uncertainties of the measured sizes.

The discovery of the extended submm/mm bright component is distinguishing  
from almost all previously studied SMGs.
 For instance, the median source size moderately bright SMGs (with a typical flux of $\sim$2 mJy at 1300 $\mu$m or $\sim$7 mJy at 850 $\mu$m) was measured to be $\sim$0$^{\prime\prime}.4$; no source larger than $1^{\prime\prime}.2$ has been reported \citep{b25}.
High-resolution radio observations are used to derive the median source size of 0$^{\prime\prime}$.65 in SCUBA SMGs in the Lockman Hole \citep{b112}.
Recently, brighter SMGs, AzTEC1 (15 mJy in 890 $\mu$m) and GN20 \citep[27 mJy;][]{b113}, have also been observed with SMA; their source sizes were estimated as approximately $0''.3$ and $0''.6$, respectively \citep{b27}; these values are similar to those of fainter SMGs. 
However, extended submm/mm structure at high redshift is observed in HzRGs such as B3 J2330+3927 \citep{b17} and 4C 60.07 \citep{b114}.
A lensed SMG, SMM J02399-0136, is known to have an extended mm bright structure, and its corrected source size is $\sim$5$^{\prime\prime}.$ \citep{b108}.

It should be noted that uv data suggesting the presence
of extended emission can also be fitted with 
a model with multiple sources. For instance,
AzTEC11, another bright SMG, can be modeled 
by two compact components 
with a separation of $\sim$2$^{\prime\prime}$ \citep{b115}.
However, in the case of Orochi, we found no multiple
compact components in the VLA image with the beam size
of $1^{\prime\prime}.3$. Therefore, Orochi has certainly an extended component, although the size has not been precisely determined yet.


\begin{figure}
\includegraphics[width=100mm]{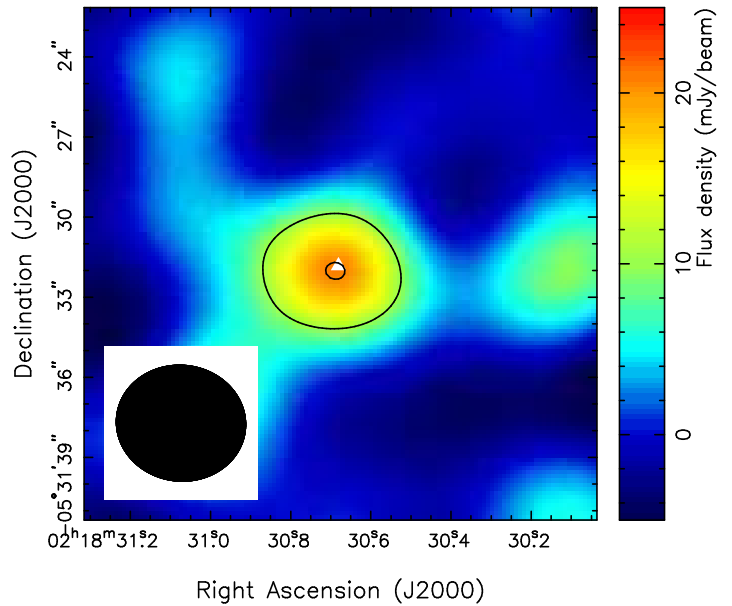}
\caption{The SMA 880 $\mu$m image of the extended component of Orochi, produced by the residual visibility data after subtracting unresolved component in the uv domain (See section 5.1).
The contour levels are 2.5, and 5 $\sigma$ (1 $\sigma=$ 4.1 mJy).
The synthesized beam is shown on the lower left. The white triangle marks the position of the compact component.
}
\label{extmap}
\end{figure}

\begin{table}
\begin{center}
\caption{Summary of submillimeter/millimeter bright components.  }
\begin{tabular}{c c c}\\ \hline \hline
 &Size& Flux density \\ \hline
\multicolumn{3}{c}{SMA 880 $\mu$m}\\ \hline
compact &unresolved& 48.6 $\pm$ 4.0 mJy \\
extended & $4^{\prime \prime}.0$ $\pm$ $1^{\prime \prime}.2$ & 42.1 $\pm$ 20.3 mJy \\ \hline
\multicolumn{3}{c}{CARMA 1300 $\mu$m}\\ \hline
compact &unresolved& 14.1 $\pm$ 3.1 mJy \\
extended & $2^{\prime \prime}.7$ $\pm$ $1^{\prime \prime}.2$ & 10.5 $\pm$ 4.3 mJy \\ \hline
\end{tabular}
\label{size}
\end{center}
\end{table}

\subsection{SED and redshift of Orochi}
We conducted SED fitting and estimated the photometric redshift of Orochi using two methods.
One involves the fitting of an SED to the optical and NIR data, and the other involves the use of submm/mm and radio data.
The former corresponds to flux from stellar emission, and the latter,  to cold dust emission.

\subsubsection{Optical and NIR photometric redshift}
The optical/NIR photometric redshift is calculated using 
the code {\it Hyperz} \citep{b76}, with 
$u_R$, $B$, $V$, $Rc$, $i^{'}$, $z^{'}$, $J$, and $K$.
The default SED template of \citet{b80},
extended using the \citet{b77} model, is adopted.
The obtained redshift is 1.39 ($\chi^2=4.7$), with the 
68\% confidence interval of $1.383-1.402$.

\begin{figure*}
\includegraphics[width=180mm]{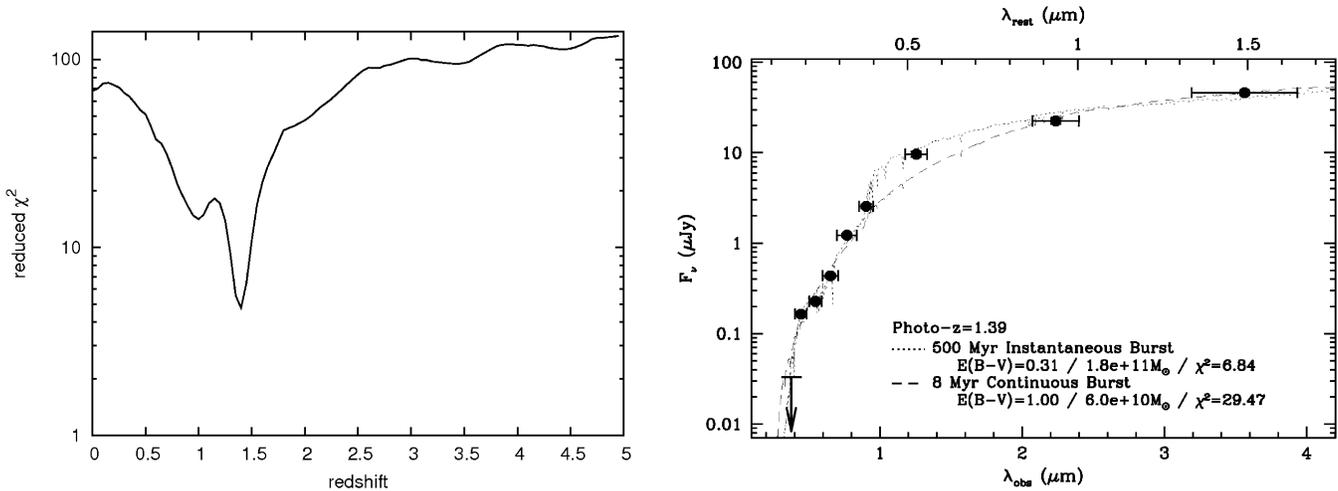}
\caption{Result of Optical/NIR photometric redshift and SED fitting of Orochi.
Left panel shows the reduced $\chi^2$ versus photometric redshift obtained by
Hyperz code.
Right panel shows the best-fit optical/NIR synthetic model spectra, by fixing
the redshift to 1.39 obtained by Hyperz.
The dotted and dashed lines show the resultant spectra assuming
instantaneous and continuous starburst, respectively.}
\label{optsed}
\end{figure*}

Next, by fixing the redshift to $z=1.39$, 
we carried out SED fitting using the 
$GALAXEV$ model \citep{b78} to 
obtain its physical properties of the galaxy.
Two star formation histories 
of instantaneous burst and 
continuous burst models are adopted, 
with the Salpeter initial mass function and 
the \citet{b79} dust extinction curve.
The results are shown in Fig. \ref{optsed}; 
it can be observed that the optical/NIR SED is well 
fitted by the 500 Myr instantaneous burst model, 
with a stellar mass of $1.8\times10^{11} M_{\odot}$
and a dust extinction of $E(B-V)=0.31$.

\subsubsection{submm/mm and radio photometric redshift}
We used a Monte Carlo photometric redshift method \citep{b116,b117} to estimate the phototmetric redshift based on submm/mm and radio data.

 A  Monte
 Carlo photometric redshift method has been developed
 to take into
 account constraining prior information such as a range
 of local SED templates, the favoured evolving luminosity function of dust enshrouded star-forming galaxies to $z\approx 2$, and the amplification of certain fields.
 We assume that the SEDs
 of SMGs are well represented by 20 SEDs
of local starbursts, ULIRGs,
 and AGN to provide FIR--radio SEDs. These SEDs cover
 a wide range of FIR luminosities ($10^{9.0} - 10^{12.3}$ $L_{\odot}$) and
 temperatures (25--65 K).
See  details in \citet{b118} for the method. 
We adopt a very conservative estimate for the amplification of
 the source ($\mu$ = 10).

We used the FIR to radio portion of the SED, i.e., the photometric data at 160 $\mu$m (Spitzer; upper limit), 1000 $\mu$m (Z-Spec), 1100 $\mu$m (AzTEC/ASTE), 1500 $\mu$m (Z-Spec) , 20 cm (VLA) and 50 cm (GMRT), for the derivation of the photometric redshift. 
These three photometric data at submm/mm wavelength were selected as representative s of the Rayleigh-Jeans slope.

The derived redshift probability distribution is shown in Fig. 9.
The estimated redshift is $3.4^{+ 0.7}_{- 0.7}$.  
The peak of the probability density distribution is very stable to changes in magnification.
Solution were derived FIR amplification factors ranging from 1 (no magnification) to 50, finding peak values between 3.3 and 3.5.
As the magnification increases, we find a slight transfer of the probability density towards higher redshift values, which displaces the 68 percent confidence interval up to 3.4$^{+0.7}_{-0.5}$.

\begin{figure*}
\includegraphics[width=180mm]{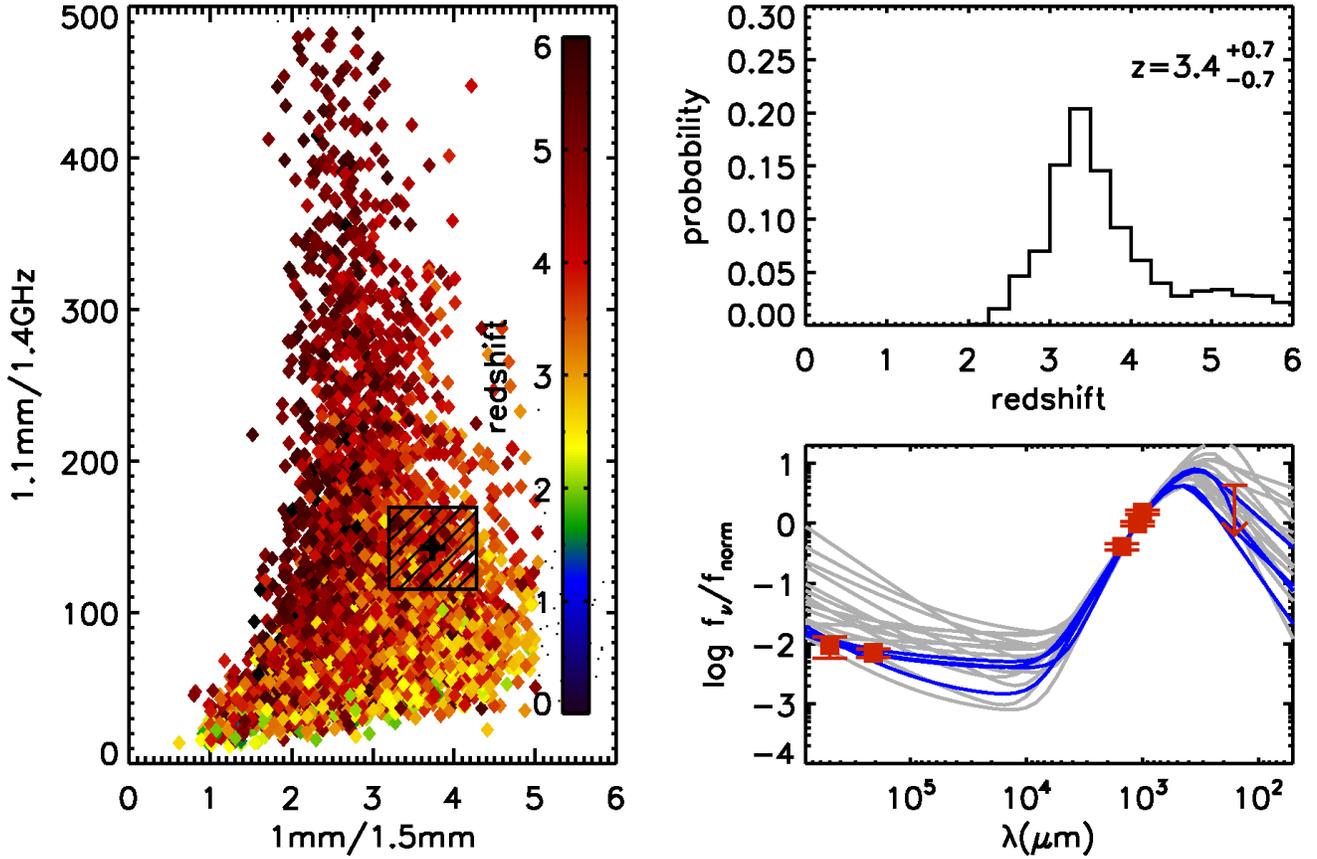}
\caption{The left panel shows the colour-colour-redshif plot for Orochi. The flux ratios of the mock galaxies are  represented as diamonds,  and their redshifts are colour-coded according to the scale shown in the panel on the right. The cross represents the measured colours of Orochi, and the dashed box shows the 1 $\sigma$ uncertanity in each colour.
The top right panel shows estimated redshift probability distribution of Orochi.
In the bottom right panel, the observed SED of Orochi normalised to the flux densities at 1000 $\mu$m is shown as squares and arrows. The arrow indicates 3 $\sigma$ upper limits. The squares denote the detection at a level $\geq$ 3 $\sigma$, with 1 $\sigma$ error bars.
The template SEDs (lines) are redshifted to $z$ = 3.4. 
The template SEDs at this redshift compatible within
3 $\sigma$ error bars with the SED of Orochi are displayed as blue lines.
The photometric data are shown at 1000, 1100, 1500 $\mu$m, 20 and 50 cm, and The upper limit is shown at 160 $\mu$m.}
\label{radiosub}
\end{figure*}

\subsubsection{$z$ $\sim$1.4 or $\sim$3.4 ?}
We now discuss  the inconsistency between the two  photometric redshift results.
The optical/NIR photometric redshift suggests $z$ $\sim$ 1.4, while the submm/mm and radio photometric redshift suggests $z$ $\sim 3.4$.
Next, we show  that this apparent discrepancy can be understood by observing
an optically dark SMG lying at $z \sim 3.4$ behind a foreground galaxy located around $z \sim 1.4$, visible at optical/NIR wavelengths.

The optical/NIR-based photometric redshift  of $\sim$ 1.4 seems to be robust
 because we obtain nearly continuous detections in optical/NIR, i.e., $\sim $0.4 $\mu$m-2.2 $\mu$m; its SED shows a clear break feature at around 1.2 $\mu$m, which is very likely to be a 4000 \AA \ break around $z \sim 1.4$.
Although the  IRAC colours of Orochi, log$(S_{5.8 {\rm \mu m}}/S_{3.6 {\rm \mu m}})=0.21$ and log$(S_{8.0 {\rm \mu m}}/S_{4.5 {\rm \mu m}})=0.28$, suggest the existence of an obscured  AGN \citep{b81},
such a sharp 4000 $\AA$ break can be reproduced solely by stellar spectra, (i.e., it is very difficult to produce a clear break with AGN models). 
Therefore, we suggest that the observed  optical/NIR spectrum is dominated by a stellar component; Further, the possible  4000  \AA \  break feature reliably constrains the redshift. The resultant reduced $\chi^2$ distribution shows a sharp peak at $z$ $\sim$ 1.4.

On the other hand, the submm/mm and radio photometric redshift is also fairly well constrained with almost continuous coverage from 880 $\mu$m to 1500 $\mu$m combined with radio and MIPS upper limits.
Although it is possible to fit the observed SED with an SMG at $z=1.4$, it requires an SED template with a very low dust temperature, $T_{\rm dust} \sim$ 20 K, which is nearly the lowest measured dust temperature distribution among SMGs \citep[e.g.,][]{b21}. Refer to  the next section for a detailed discussion of SED modeling with a grey-body.

  Another constraint on the redshift of the gas/dust component comes from Z-Spec observations. 
In general, the line-to-continuum ration (L/C ratios) are expected to be smaller for higher J transitions because the continuum flux increases with $\nu^{3-4}$.
On the other hand, the CO line flux increases with $J^2$ or $\nu^2$ for thermalized conditions, and since high-$J$ lines are often subthermally excited, they are much weaker than in a thermalized case.

As no  emission line features are detected for the intense continuum emission, the upper limit on the L/C ratio is 0.1--0.3 (2 $\sigma$) accross the Z-Spec band. 
This value is  low if we assume a redshift of $\sim$ 1.4, where CO($J$=4-3), CO($J$=5-4), and CO($J$=6-5) whould be expected to be in the Z-Spec band.

We collected the L/C ratios from the measured CO fluxes and adjacent continuum emission using interferometers in the literature and we found that L/C ratios are typically in the range of 3--20 for high-z quasars \citep{b92,b91,b29,b94,b95}, radio galaxies \citep{b96}, SMGs \citep{b90,b25}\footnote{There are no direct descriptions of the continuum fluxes, and L/C ratios are estimated from the CO($J$=4-3) spectrum in these papers.}, and a local ULIRG \citep{b97}. This indicates that we have already achieved  sufficient line sensitivity for the observed continuum flux level if Orochi is lying at around $z \sim 1.4$.

On the other hand, the observed low L/C ratio is acceptable if Orochi is at $z \sim 3.4$; this is because the observed $J$ lines such as CO($J$=9-8) to CO($J$=13-12) in the Z-Spec band are considered to be of a much higher order.
Smaller L/C ratios around unity have been reported for CO($J$=9-8) in APM J08279+5255  \citep{b92,b91} and the Cloverleaf \citep{b94}. 
It should be emphasized that these two quasars show exceptionally highly excited CO lines up to  CO($J$=10-9) level \citep{b91}; the measured CO SED suggests that the CO fluxes are consistent with the thermalized condition as they  are expected to increase with $\nu^2$ up to $J$ $\sim$ 10. 
Other typical high-z quasars as well as starburst cores of local IR bright galaxies like M82 and NGC 253 become subthermal at around $J \sim$ 4 -- 6; a simple LVG model calculation suggests that the expected CO($J$=9-8) flux in a typical starburst core (e.g., gas density $n_{\rm H_2} = 10^4$ cm$^{-3}$ and kinetic temperature $T_{\rm kin} = 60$ K) will be smaller than the thermalized-case flux by a factor of 20 or more; i.e., it is natural to expect that the L/C ratio for these high-$J$ lines are much smaller than unity, as  observed in Orochi with Z-Spec.

Here, we briefly comment on the constraint on redshift from [CII] emission ($\nu_{\rm rest}$ = 1.9019 THz); this line can be in the Z-Spec band if $5.2<z<9.5$.
 The upper limit to the [CII] line luminosity is $L_{\rm [CII]} < 3.6 \times 10^9 L_\odot$ (2 $\sigma$), if it is at $z=5.5$ with a line width of 780 km s$^{-1}$, which is a median value for SMGs \citep[e.g.,][]{b90}.
This results in a strict upper limit on the [CII] to FIR luminosity ratio, $L_{\rm [CII]}/L_{\rm FIR} < 3.6 \times 10^{-5}$ (2 $\sigma$), which is already much smaller than the observed $L_{\rm [CII]}/L_{\rm FIR}$ ratios for high-z quasars/SMGs, i.e., $\sim 10^{-4} - 10^{-3}$ \citep{b31,b93,b87}. 
We, therefore, suggest that the redshift of Orochi would be less than $\sim$ 5.

In summary,  we  suggest that the mm/submm bright component of Orochi is likely to be  at $z$ $\sim$ 3.4, whereas  the optical/NIR counterpart candidate with a photometric redshift of $\sim$ 1.4 is a foreground galaxy with an old stellar population. 
Although it is difficult to ignore the possibility  that both systems, the i.e., submm-to-radio bright component and its optical/NIR counterpart, are lying at the same redshift (i.e., at $z \sim 1.4$), this possibility requires an exceptionally low dust temperature for SMGs ($T_{\rm dust} \sim 20$ K)  as well as a low-excitation molecular gas for which the mid-$J$ lines such as CO($J$=4-3) are not well excited. 

In the following sections, we estimate physical quantities for two cases --- $z \sim 3.4$ and $z \sim 1.4$ --- because it is difficult to completely reject the latter possibility at this moment, even though the former seems more probable.

\subsubsection{Dust SED  model using a grey-body}
We model  SEDs from FIR to radio wavelengths using grey bodies.
To draw SEDs at radio wavelengths, we assume the radio-FIR correlation \citep[e.g.,][]{b16,b129}.
This correlation is described as the  q-value
\footnote{The q-value is defined as
$
q \propto \log(L_{\rm FIR}) - \log(I_{1.4\mathrm{GHz}}),
$
where $I_{1.4\mathrm{\rm GHz}}$ is a rest-frame 1.4 GHz luminosity density and $L_{\rm FIR}$ is defined as the 42.5--122.5 $\mu$m FIR luminosity.
This relation corresponds to the median q-value, $<q> \sim 2.3$, with a rms scatter of $\leq$0.2 in  nearby normal galaxies \citep{b16}
and 2.34 $\pm$ 0.01 in local IRAS galaxies \citep{b103}.
On the other hand, we  use the total 8--1000 $\mu$m IR ($L_{\rm IR}$) as an indicator of dust emission.
In this case, the q-value is redefined as $q_{\rm TIR} \propto \log(L_{\rm IR}) - \log(I_{1.4 {\rm GHz}})$.
The value of $q_{\rm TIR}$ is 2.64 $\pm$ 0.02 in local galaxies, which is consistent with the above q-value \citep{b119}. }.
It is known that the mean of $q_{\rm TIR}$
for SMGs is 2.32 with a scatter of 0.34 \citep{b7}. 
We use this typical $q_{\rm TIR}$ of known SMGs for extrapolating the radio spectrum.
Moreover,
$L_{\rm  IR}$ is derived from the SED of a model grey-body.

We also derived the radio emissivity index $\alpha$ ($S \propto \nu^{\alpha}$) of Orochi.
Orochi has two photometric data at radio wavelengths, 20 cm (VLA) and 50 cm (GMRT).
From these data, the radio emissivity index $\alpha$ of Orochi is derived as $-0.33^{+0.29}_{-0.32}$. 
Comparing $\alpha$ of local star-forming galaxies \citep[$\sim$ -0.7 to -0.8;][]{b16}, and SMGs at $z$ $\sim$ 2 \citep[$-0.75$ $\pm$ 0.06;][]{b124}, the measured $\alpha$ in Orochi is found to be rather closer to a pure thermal radio spectrum. 
If Orochi is located at $z >$ 3, the $\alpha$ of Orochi seems to be consistent with predictions 
that non-thermal radio continuum emission will be suppressed at
high $z$ ($z >$ 3 or so) due to the increased energy losses from the inverse Compton
 scattering off in
the cosmic microwave background to the cosmic ray electrons \citep{b120}.

\begin{figure}
\includegraphics[width=90mm]{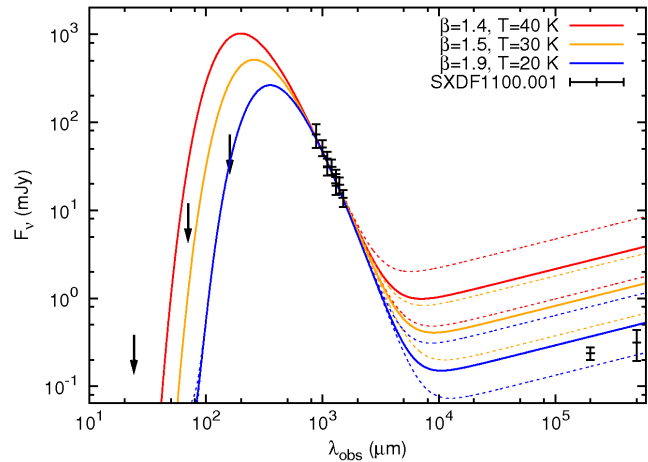}
\caption{The SED of Orochi and SED fitting results in FIR-radio wavelengths, assuming $z$=1.4. We calculated grey-bodies for  $T_{dust}=$ 20, 30, and 40 K. The $\beta$ for each grey-body is estimated by fitting the Z-spec continuum data at 1000--1500 $\mu$m. The dashed lines show the scatter of the q-value. The upper limits are shown at 24, 70, and 160 $\mu$m, and the photometric data are shown at 880, 1000, 1100, 1200, 1300, 1400, 1500 $\mu$m, 20 and 50 cm. We find that the dust temperature of Orochi must be as low as 20 K  if we assume $z$ $\sim$ 1.4. }
\label{SED}
\end{figure}

For the models based on the optical/NIR photometric  redshift of 1.4, the dust emissivities were derived as  $\beta=$ 1.4, 1.5, and 1.9 for $T_{\rm dust}=$ 40, 30, and 20 K, respectively, 
 by fitting the Z-Spec continuum data at 1000--1500 $\mu$m.
We find that a low dust temperature, 20 K, is required if a photometric redshift of 1.4 is assumed.
 Possibilities of higher dust temperatures such as 30 K and 40 K are clearly eliminated because the upper limits at 24, 70, and 160 $\mu$m bands  tight constraints on the dust temperature. 
The radio flux density also seems to be consistent with a model with $T_{\rm dust}$ of 20 K. 
Again, higher dust temperatures such as 30 K and 40 K are not acceptable because at such high dust temperatures,  $L_{\rm IR}$ is overluminous in terms of radio luminosity, as clearly shown in Fig. \ref{SED}.

On the other hand, for  models based on the submm/mm-to-radio photometric redshift of $\sim$3.4, 
we find that 
the observed submm/mm to radio SED can be well explained by  the dust temperature of $\sim$ 30 K, as shown in Fig. 12.
This is consistent with the typical dust temperature of SMGs  \citep[e.g.,][]{b21}.

\begin{figure}
\includegraphics[width=90mm]{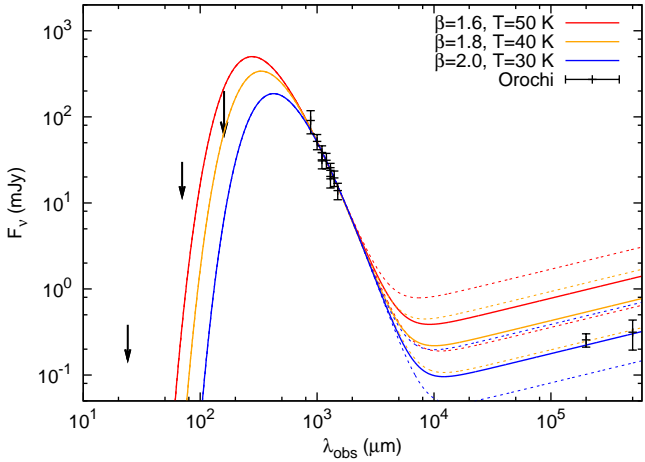}
\caption{The SED of Orochi and SED fitting results in FIR-radio wavelengths, assuming $z$=3.4. We calculated grey-bodies for  $T_{dust}=$ 30, 40, and 50 K. The $\beta$ for each grey-body is estimated by fitting the Z-spec continuum data at 1000--1500 $\mu$m. The dashed lines show the scatter of the q-value. The upper limits are shown at 24, 70, and 160 $\mu$m, and the photometric data are shown at 880, 1000, 1100, 1200, 1300, 1400, 1500 $\mu$m, 20 and 50 cm. We find that the dust temperature of Orochi seems to be as low as 30 K  if we assume $z$ $\sim$ 3.4.}
\label{SED3}
\end{figure}

\subsection{A possible positional offset between a submm-radio bright component and an optical/NIR counterpart candidate}
Here, we present a detailed comparison of the multi-wavelength images of Orochi. Fig.~13 displays the peak positions of CARMA, SMA, and VLA sources listed in Table 3 superposed on the Subaru $z^{\prime}$-band image. We find a possible positional offset between the optical peak and the submm/mm/radio peaks; its separation is approximetely $0''.1$--$0''.7$. 
Although these separations are smaller than the beam sizes of the observations, the evaluated combined-astrometry errors listed in Table 3 are  smaller than the claimed offset, especially in declination. 
Furthermore, all three observations (i.e., CARMA, SMA, and VLA) show a similar tendency, i.e., the peaks
of these wavelengths are located at the north west part with respect to the $z^{\prime}$-band peak, which is also  associated with the $K$ band peak. 
These facts  suggest the possible existence of a spatial offset between the peaks in the optical/NIR and the submm/mm/radio wavelengths. 

As discussed in section 5.2.3, the optical/NIR bright counterpart candidate can be a foreground source of Orochi.
In this case, the extended bright structure observed in the optical and NIR bands, i.e., the north-west extension from the optical/NIR peak positions around the  SMA, CARMA, and VLA is probably a true  optical counterpart of Orochi and is blended with the foreground source.

In order to confirm this possibility, we need higher angular resolution images for all wavelengths. 
Future ALMA observations combined with HST or ground-based AO observations will be conducted for addressing this issue.

\begin{figure}
\includegraphics[width=90mm]{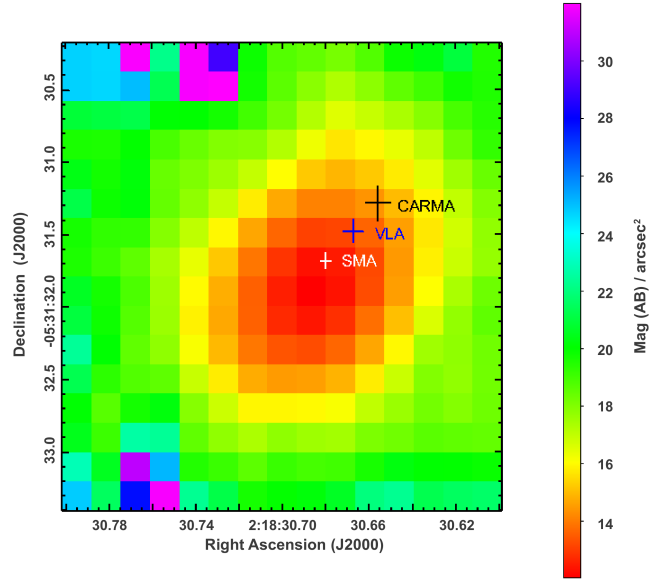}
\caption{Radio/mm/submm peak positions superposed on the $z$-image of Orochi.
The white, black,  and blue  crosses (expressing the position uncertainties)  correspond to the SMA, CARMA, VLA peak positions, respectively.
The unit of the background $z^{\prime}$-band image is Mag (AB) per ${\rm arcsec}^2$.}
\label{zure}
\end{figure}

\subsection{The Origin of 1100 $\mu$m Flux in Orochi}

The observed 880 $\mu$m flux of Orochi, $\sim$ 91  mJy, makes this one of the brightest SMGs known to date. For instance, it is much brighter than the brightest SMGs revealed by SHADES \citep[$\sim$ 10-15 mJy at 850 $\mu$m;][]{b52}
and brighter or comparable to the  known lensed submillimeter bright high-z sources such as the Cloverleaf
\citep[58.8 $\pm$ 8.1 mJy at 850 $\mu$m, with a lens magnification of 11;][]{b39,b40}, APM08279+5225 
\citep[84 $\pm$ 3 mJy at 850 $\mu$m, with a lens magnification of 7;][]{b39,b41}
 and SMMJ2135-0102 \citep[106.0 $\pm$ 7.0 mJy at 870 $\mu$m, with a lens magnification of 32;][]{b74}.
 A similar bright SMG has also been uncoverd by MAMBO on IRAM 30 m observations \citep[30 $\pm$ 2 mJy at 1200 $\mu$m;][]{b125}.
SPT surveys detected 47 dust-dominated mm sources above 10 mJy at 1.4 mm in 87 deg$^2$, and 20 of 47 do not have counterparts at low-$z$ universe \citep{b71}.
If all of 20 are indeed SMGs located at high-$z$ universe, a surface density of these ultra-bright SMGs is $\sim$ 0.23 sources per 1 deg$^{2}$.
Considering we found only 1 ultra-bright SMG in about  0.33 deg$^2$ , which is surveyed area in SXDF, it is consistent that Orochi belongs to the same population with SPT sources.

 It is possible  that Orochi is gravitationally lensed, because many of the ultra-bright SMGs are often strongly amplified by a foreground massive cluster of galaxies 
\citep[e.g.,][]{b5,b83,b74,b87}.
The recent model analysis for submm/mm count shows that these ultra bright SMGs are likely to be gravitationally lensed by foreground massive halos \citep{b123,b127}.
Although we found no catalogued foreground clusters around Orochi,
analysis of photometric redshifts of optically visible objects around Orochi suggest that  there are cluster candidates at $z$ $\sim$ 0.4 and 1.4, and they may be partially responsible for lensing.

It is also possible  that this SMG is gravitationally lensed  by other foreground galaxies near Orochi. Theoretically, when a galaxy causes a lensing effect on another galaxy, the former galaxy is located at around less than  1 arcsec from the latter. In the case of Orochi, the optical/NIR bright foreground galaxy at $z$ $\sim$ 1.4 can be a candidate for the lensing galaxy. The estimated stellar mass of the optical/NIR bright source by SED fitting is $\sim$ $10^{11} \ {\rm M_{\odot}}$; therefore, it is sufficiently massive to cause gravitational lensing.
In fact, some of 500$\mu$m-selected ultra bright SMGs are indeed gravitationally lensed by foreground galaxies \citep{b126}.


Nevertheless, we note that 
 there is no clear evidence for  strong lensing in existing high-resolution optical/infrared images, for instance, the presence of multiple and/or distorted structures. 
This situation is  different  from that of SMM J2135-0102, another ultra-bright SMG, where distorted/elongated images can be clearly observed even in the resolution of the IRAC bands, although the noise level of the IRAC bands in  SXDF is at least 10 times better than that of SMM J2135-0102 \citep{b74}.

On the other hand, the situation of Orochi is similar to that of HDF850.1  \citep[7.0 $\pm$ 0.4 mJy at 850 $\mu$m;][]{b49}. The photometric redshift of HDF850.1 is $z$ = 4.1$^{+ 0.6}_{- 0.5}$ \citep{b122} and it is indicated that HDF850.1 is gravitationally lensed by an elliptical galaxy at $z=$1.22 about 1 arcsec away from HDF850.1 \citep{b49,b121,b122}.
HDF850.1 also has no clear counterpart image in the mid- and near-infrared deep images.


The elevated mm/submm fluxes can be  from non-thermal synchrotron emission if they are powered by a radio-loud AGN. In fact, wide-field mm surveys often uncover such non-thermal sources \citep[e.g.,][]{b70,b71}.
 In the case of Orochi, however,  we conclude that  the submillimeter flux is dominated by thermal dust emission heated by massive stars and not by a radio-loud  AGN because the submm-to-radio SED can be well reproduced by the known $L_{\rm IR}$-to-radio correlation observed in  previously studied SMGs and local starburst galaxies, as shown in section 5.2.4.
Absence of the time variability in the submm/mm fluxes of Orochi, i.e., in the 1100 $\mathrm{\mu m}$ flux measured by AzTEC in November/December 2008, the 1300 $\mathrm{\mu m}$ flux by CARMA in August 2009, the 880 $\mathrm{\mu m}$ flux by SMA in December 2009, and the Z-Spec data in November 2009,
may also suggest  the thermal origin of the elevated flux of Orochi.

\subsection{Estimation of physical properties of Orochi}
We estimate the physical quantities of Orochi, i.e., $L_{\rm IR}$, dust mass ($M_d$) and gas mass ($M_{\rm gas}$), by adopting
two possible redshifts and models  in section 5.2:
(1) $z=$ 1.4, for which the SED model of the grey-body with $T_d$ = 20 K and $\beta$ = 1.9, and
(2) $z=$ 3.4, for which we used an averaged SED template of the 4 best fitted SEDs (the SED templates of NGC3227, NGC7771, IRAS05189-2524, and IRAS12112+0305), providing $T_d$ =35 K and $\beta$ = 2.0.
Here, we simply show the physical quantities of Orochi using the apparent luminosity, i.e., $L_{\rm IR} =$ $\mu L^{*}_{\rm IR}$, where $L^{*}_{\rm IR}$ is the intrinsic luminosity and $\mu$ is the magnification factor, because we have no constraint on the magnification factor (if any) at this moment. Furthermore, here, we do not distinguish between the compact and extended components.

We estimated SFR using the equation below.
\begin{equation}
SFR (\mathrm{M_{\odot}yr^{-1}}) = 4.5 \times 10^{-44} L_{\rm IR} (\mathrm{ergs \ s^{-1}})  \label{eq:sfr}
\end{equation}
\citep{b8}.
$M_d$ is derived as
\begin{equation}
M_d =\frac{ S_{obs}D^2_{L}}{\left(1 + z \right) \kappa_d (\nu_{rest}) B(\nu_{rest}, T_{dust}) },
\end{equation}
where $S_{obs}$ is the observed flux density. In this paper, the flux density is at 1100 $\mu$m, $\nu_{rest}$ is the rest-frame frequency; $\kappa_d(\nu_{rest})$, the dust mass absorption coefficient; and $B(\nu_{rest}, T_{dust})$,  the Planck  function \citep{b18}.
We assume that the absorption coefficient varies as $\kappa_d \propto \nu^{\beta}$.
We adopt $\kappa_d (125 \ \mathrm{\mu m}) = 2.64 \pm 0.29 \ \mathrm{m^2 kg^{-1}}$, the average value of various studies \citep{b20}.
The gas mass is then  derived using a gas-to-dust mass ratio of 54, which is an average value for SMGs in \citet{b21}.

The derived quantities are listed in Table \ref{phys}, along with the gas consumption time scale, which is defined as the ratio of $M_{\rm gas}$/SFR or 1/SEF, where SEF refers to the star formation efficiency. 
However it should be noted that many of these results are rough estimates based on median relations, assumption about the SED.

Table \ref{phys} shows 
that if $z$ $\sim$ 3.4, Orochi will consume 
its gas reservoir within a short time scale ($\sim 3 \times 10^{7}$ yr), which is indeed comparable
to those in extreme starbursts like the hearts of local ULIRGs \citep[e.g.,][]{b8}.
On the other hand, if this source has a lower $z$ of $\sim$ 1.4, Orochi has a fairly
long gas consumption time scale of $\sim$1 $\times 10^{9}$ yr, which is almost comparable to that of star-forming clouds in the disks of spiral galaxies in the local universe, and hence, rather unlikely.

\begin{table*}
\begin{center}
\caption{ Estimated IR luminosity, dust mass, gas mass, star formation rate, and inverse of star formation efficiency (corresponding to a gas consumption time scale)  for two possible  redshifts, $z$ = 1.4 (model 1) and 3.4 (model 2), which are esitimated in section 5.2.4.
Here, we describe the observed quantities as $\mu L^{*}_{\rm IR}$, $\mu M^{*}_{\rm d}$,  $\mu M^{*}_{\rm gas}$, and $\mu$SFR$^{*}$
 , where $\mu$  indicates the magnification factor. Note that 1/SFE
is independent of magnification.}
\begin{tabular}{c c c c c c c c c}\\ \hline \hline
model & $z$ & $T_d$ & $\beta$ & $\mu L^{*}_{\rm IR}$                            &$\mu M^{*}_{\rm d}$                                &$\mu M^{*}_{\rm gas}$                              &  $\mu$SFR$^{*}$                                  & 1/SFE \\ \hline
(1)   & 1.4 & 20 K  & 1.9      & $5.1 \times 10^{12} \  L_{\odot}$  &$1.7 \times 10^{10} \ M_{\odot}$ &$9.1 \times 10^{11} \ M_{\odot}$ & $870 \ M_{\odot}  {\rm yr^{-1}}$ &$1.0 \times 10^{9} \ \mathrm{yr}$ \\
(2)   & 3.4 & 35 K  & 2.0      & $6.3 \times 10^{13} \  L_{\odot}$  &$5.8 \times 10^{9} \  M_{\odot}$ &$3.1 \times 10^{11} \ M_{\odot}$ & $11000 \ M_{\odot}  {\rm yr^{-1}}$ &$2.8 \times 10^{7} \ \mathrm{yr}$ \\
\hline
\end{tabular}
\label{phys}
\end{center}
\end{table*}

\subsection{Starburst form of Orochi}

One of the remaining key questions of Orochi is its form of starbursts: What type of star formation or starburst  produces such an SFR elevated to $\sim$ 870 $M_\odot$ yr$^{-1}$ (if $z \sim 1.4$) or 11000 $M_\odot$ yr$^{-1}$ (if $z \sim 3.4$)? Recent high angular resolution studies of high-z quasars and SMGs suggest that Eddington-limited maximal starbursts often occur at the heart of these extreme objects \citep{b9,b42}. In these galaxies, the peak SFR and $L_{\rm IR}$ surface densities reach up to $\Sigma_{\rm SFR} \sim 10^{3}$ $M_\odot$ yr$^{-1}$ kpc$^{-2}$and $\Sigma_{L_{\rm IR}} \sim 10^{13}$ $L_{\odot}$ kpc$^{-2}$, respectively; these values are close to the theoretical maximum value  imposed by the Eddington limit \citep{b43,b44}. Therefore, it is intriguing to address whether the SFR/$L_{\rm FIR}$ surface densities at the center of Orochi are close to the maximum starburst condition or not.

We estimate the surface gas mass density ($\Sigma_{gas}$), SFR surface density ($\Sigma_{SFR}$) and $L_{\rm IR}$  surface density ($\Sigma_{L_{\rm IR}}$) for 
these two components (Table 7), respectively.

We assume that the source size of the unresolved compact component is $0^{\prime \prime}.4$, which is the median size of the submillimeter bright region of SMGs \citep{b24,b25}. 
The derived surface densities for two possible photometric redshifts, $\sim$ 1.4 and $\sim$ 3.4, are listed in Table \ref{phys2}.

The $\Sigma_{gas}$ and other derived densities of the extend component are similar to or  higher than those of the star-forming disk regions in local spiral galaxies \citep[e.g.,][]{b45,b26}.

This indicates that the extended component of Orochi is a large lump of relatively moderate starburst regions, 
although we should note that the extended structures can be a lensed image. We then need to estimate the magnification factor to discuss the quantitative properties of star formation. 

On the other hand, the derived parameters of the compact component
are rather close to those of ULIRGs in the local universe, i.e.,  the predicted maximum limit \citep[e.g.,][]{b42}, although it still depends on the assumed source size and magnification factor (if any).

\begin{table*}
\begin{center}
\caption{ Estimated surface density of molecular gas mass, SFR, and $L_{\rm IR}$ ( $\Sigma_{\rm gas}$, $ \Sigma_{\rm SFR}$, and $\Sigma_{L_{\rm IR}}$, respectively) for two components revealed by SMA, i.e., spatially unresolved, compact component (the size was assumed to be $0''.4$) and spacially extended component ($4''.0$). We adopted  a scale of 8.432 kpc/arcsec for $z = 1.4$ and  7.396 kpc/arcsec for $z$ = 3.4. Adopted SED models, dust temperatures, and parameters are the same as Table \ref{phys}.  }
\begin{tabular}{c c c c c c c c}\\ \hline \hline
Model & $z$ & $T_{\rm dust}$ &Component & Size                                 &      $\Sigma_{\rm gas}$                          & $ \Sigma_{\rm SFR} $                                              & $\Sigma_{L_{\rm IR}}$ \\ \hline
\multirow{2}{*}{(1)} &\multirow{2}{*}{1.4} & \multirow{2}{*}{20 K}  & compact  & $0^{\prime \prime}.4$ (3.4 kpc)     &  $5.4 \times 10^{10} \ M_{\odot} {\rm kpc^{-2}}$ & 5.2 $\times 10^1 \ M_{\odot} \mathrm{yr^{-1} kpc^{-2}}$ & $ 3.0 \times 10^{11} \  L_{\odot} \mathrm{kpc^{-2}}$ \\
     &   &        & extended  & $4^{\prime \prime}.0$ (33.7 kpc)     &  $4.1 \times 10^{8} \ M_{\odot} {\rm kpc^{-2}}$ & 3.9 $\times 10^{-1} \ M_{\odot} \mathrm{yr^{-1} kpc^{-2}}$ & $ 2.3 \times 10^{9} \  L_{\odot} \mathrm{kpc^{-2}}$ \\ \hline
\multirow{2}{*}{(2)} &\multirow{2}{*}{3.4} & \multirow{2}{*}{35 K}  & compact  & $0^{\prime \prime}.4$ (3.0 kpc)     &  $2.4 \times 10^{10} \ M_{\odot} {\rm kpc^{-2}}$ & 8.3 $\times 10^2 \ M_{\odot} \mathrm{yr^{-1} kpc^{-2}}$ & $ 4.8 \times 10^{12} \  L_{\odot} \mathrm{kpc^{-2}}$ \\
     &   &        & extended  & $4^{\prime \prime}.0$ (29.6 kpc)     &  $6.7 \times 10^{7} \ M_{\odot} {\rm kpc^{-2}}$ & 7.4 $\times 10^0 \ M_{\odot} \mathrm{yr^{-1} kpc^{-2}}$ & $ 4.3 \times 10^{10} \  L_{\odot} \mathrm{kpc^{-2}}$ \\ \hline
\end{tabular}
\label{phys2}
\end{center}
\end{table*}

\section*{Summary}
An ultra-bright SMG, Orochi, 
has been detected using with the AzTEC on ASTE. Subsequent CARMA, SMA, and Z-Spec observations confirm the AzTEC/ASTE detection of Orochi.
The major findings and conclusions are summarized as follows.

\begin{enumerate}

\item We discovered a 37.27 $\pm$ 0.65 mJy source at $\lambda$ = 1100 $\mu$m in SXDF using AzTEC mounted on ASTE. 

\item CARMA $\lambda=$1300 $\mu$m and SMA $\lambda=$880 $\mu$m observations successfully confirmed the AzTEC/ASTE detection of Orochi.
The peak positions of the CARMA/SMA sources coincide with the AzTEC/ASTE peak position. In addition, these flux densities
from 1300 to 880 $\mu$m are consistent with a single  SED for $\beta=1 \sim 2$. 
The 880 $\mu$m flux density of Orochi is $\sim$91 mJy,
which makes it one of the brightest SMGs after SMMJ2135-0102 ($\sim$106 mJy at 870 $\mu$m).

\item The CSO 10 m telescope equipped with Z-Spec was used 
to conduct a blind search for
redshifted molecular/atomic lines and continuum emission
in the 190--308 GHz band. The measured continuum flux is consistent to the SMA/AzTEC/CARMA measurements, although no significant emission/absorption line features were found. The derived upper limit to the line-to-continuum flux ratio (L/C ratio) was 0.1--0.3 (2 $\sigma$) across the Z-Spec band. 

\item We find that Orochi is spatially resolved and has two components,
i.e., an extended structure (FWHM of  $\sim 4^{\prime\prime}$) and a compact unresolved
one, based on the analysis of the visibility amplitude as a function of the projected
baseline length in both SMA and CARMA data. Approximately half of the total millimeter/submillimeter flux arises from the extended component. The discovery of
the extended submm/mm bright component is distinguishing because such an extended
bright structure has not been observed in previously studied normal SMGs, which show a median
source size of $0^{\prime\prime}.4$. 

\item Multi-wavelength counterparts of Orochi have been identified near the peak position of CARMA/SMA using optical (Subaru), NIR (UKIRT; Spitzer),  MIR(Spitzer) and radio (VLA and GMRT) images.

\item A robust photometric redshift of 1.4  was derived  using optical/NIR data and the code {\it Hyperz} based on a detection of possible 4000 \AA \ break feature. On the other hand, a photometric redshift using submm/mm and radio data  suggests $z$ $\sim$ 3.4, eliminating the possibility of $z<3$. A fairly low dust temperature, $T_{\rm dust} \sim 20$ K, is required if $z=1.4$, whereas $T_{\rm dust}$ of 30--50 K, typical for SMGs,  can explain the observed FIR to radio SED for a redshift range of 3--5.

\item The discrepancy in photometric redshifts can be understood for an optically dark SMG lying at $z \sim $ 3.4 with a foreground galaxy around $z \sim$ 1.4. Indeed, we find a positional offset of $\sim$0$^{\prime\prime}.2$-0$^{\prime\prime}.7$  between sub/mm/radio peaks and and optical peak, implying a possible coincidental overlap of two objects along the line of sight.

\item If Orochi is located at $z \sim$ 3.4, higher order $J$-lines  such as CO($J=9-8$) to CO($J=13-12$) fall into Z-Spec band.
However we did not achieve any significant detection of lines.
The upper limit on L/C $\sim$ 0.1-0.3 in Z-Spec band is expected for these very high $J$ lines.

\item If the millimeter/submillimeter bright component of Orochi is indeed lying
at $z \sim 3.4$, the deduced apparent FIR luminosity ($L_{\rm IR}$) and star formation rate (SFR) are $L_{\rm IR} \sim 6 \times 10^{13} \ L_{\odot}$ and SFR $\sim 11000 \ \mathrm{M_{\odot}yr^{-1}}$, respectively, if the huge $L_{\rm IR}$ is originated from a massive starburst. The apparent surface densities of $L_{\rm IR}$ and SFR, $\Sigma_{\rm gas}$ and $\Sigma_{L_{\rm IR}}$, of the unresolved compact component are similar to those of local ULIRGs’ cores, i.e., close
to the theoretically expected maxmum value imposed by the Eddington limit, although a constraint is required on the magnification factor (if any) to yield further quantitative discussions.

\end{enumerate}
\section*{Acknowledgments}
We would  like to thank everyone
 who helped staff and support the AzTEC/ASTE 2008 operations and data calibration, including N. Ukita, M. Tashiro,  M. Uehara, S. Doyle, P. Horner, J. Cortes, J. Karakla, and G. Wallace.
The ASTE project is driven by the Nobeyama Radio Observatory (NRO), a branch of
the National Astronomical Observatory of Japan (NAOJ),
in collaboration with the University of Chile and Japanese
institutions including the University of Tokyo, Nagoya University, Osaka Prefecture University, Ibaraki University, and Hokkaido University. 
Partial observations with ASTE were  carried out remotely from Japan using NTT's GEMnet2 and its partner R\&E networks, which are based on the AccessNova collaboration of the University of Chile, NTT Laboratories, and the NAOJ. 
This study was supported in part by the MEXT Grant-in-Aid for Specially Promoted Research (No. 20001003).
The Submillimeter Array is a joint project between the Smithsonian 
Astrophysical Observatory and the Academia Sinica Institute of Astronomy 
and Astrophysics and is funded by the Smithsonian Institution and the 
Academia Sinica.
Support for CARMA construction was derived from the Gordon and Betty Moore Foundation, the Kenneth T. and Eileen L. Norris Foundation, the James S. McDonnell Foundation, the Associates of the California Institute of Technology, the University of Chicago, the states of California, Illinois, and Maryland, and the National Science Foundation. Ongoing CARMA development and operations are supported by the National Science Foundation under a cooperative agreement, and by the CARMA partner universities.

\label{lastpage}

\end{document}